\documentclass{article} 
\usepackage[T1]{fontenc}
\usepackage{iclr2027_conference,times}

\usepackage{amsmath}
\usepackage{amssymb}
\usepackage{booktabs}
\usepackage{array}
\usepackage{graphicx}
\usepackage{microtype}
\usepackage{xcolor}
\usepackage{listings}
\usepackage{enumitem}
\usepackage{xspace}
\usepackage{url}
\usepackage{hyperref}

\title{How Should Diffusion Language Models Edit Code?}

\newcommand{\hkuaff}{\ensuremath{\spadesuit}}
\newcommand{\huaweiaff}{\ensuremath{\heartsuit}}
\newcommand{\cityuaff}{\ensuremath{\clubsuit}}
\newcommand{\pkuaf}{\ensuremath{\diamondsuit}}

\author{
Xijia Tao\textsuperscript{\hkuaff,*} \quad
Ziru Liu\textsuperscript{\huaweiaff,*} \quad
Shansan Gong\textsuperscript{\hkuaff,*} \quad
Jiacheng Ye\textsuperscript{\hkuaff} \quad
Kecheng Chen\textsuperscript{\cityuaff} \\
\bfseries
Zirui Wu\textsuperscript{\pkuaf} \quad
Lin Zheng\textsuperscript{\hkuaff} \quad
Xinyu Fu\textsuperscript{\huaweiaff} \quad
Rui Liu\textsuperscript{\huaweiaff,\textdagger} \quad
Lingpeng Kong\textsuperscript{\hkuaff,\textdagger} \\
\textsuperscript{\hkuaff}The University of Hong Kong \quad
\textsuperscript{\huaweiaff}Huawei Technologies Ltd. \\
\textsuperscript{\cityuaff}City University of Hong Kong \quad
\textsuperscript{\pkuaf}Peking University \\
\textsuperscript{*}Equal contribution \quad
\textsuperscript{\textdagger}Corresponding authors
}

\newcommand{\dllm}{dLLM\xspace}

\newcommand{\strreplace}{\texttt{str\_replace}\xspace}
\newcommand{\masktok}{\texttt{[MASK]}}
\newcommand{\expandtok}{\texttt{<|expand|>}}

\iclrfinalcopy
\hypersetup{
  pdftitle={How Should Diffusion Language Models Edit Code?},
  pdfauthor={Xijia Tao, Ziru Liu, Shansan Gong, Jiacheng Ye, Kecheng Chen, Zirui Wu, Lin Zheng, Xinyu Fu, Rui Liu, Lingpeng Kong}
}

\begin{document}

\maketitle
\fancyhead{}
\suppressfloats[t]

\begin{abstract}
Code editing requires a model to decide \emph{where} to make changes, generate the new content, and preserve everything else.
We study how masked diffusion language models divide these responsibilities across four editing interfaces: whole-file rewriting, search-and-replace, locate-then-infill, and token-level editing.
Experiments on CanItEdit reveal a \emph{composition gap}: diffusion models can generate coordinated changes when the correct edit locations are supplied, but much of this capability is lost when those locations must be predicted.
Access to the intact original code helps the model fill multiple edit regions, yet does not resolve the difficulty of selecting those regions.
By varying the editable regions while holding the generation model and decoding procedure fixed, we identify two distinct requirements for successful editing: covering every required change and placing precise boundaries around it.
Missing a required region prevents the corresponding change, while widening regions to ensure coverage can sharply reduce success by requiring unchanged code to be regenerated.
A sentence-level Wiki editing probe shows the same qualitative gap between supplied and predicted locations beyond code.
These findings show why strong infilling capability alone does not ensure reliable editing: the interface must expose all required changes while limiting regeneration of unchanged code.
\end{abstract}

\section{Introduction}
AI coding assistants often work on programs that already exist. A request to fix a bug, extend an API, or update its callers requires coordinated changes while preserving behavior outside the request. Instructional code editing makes this requirement explicit: the model receives an instruction and an intact source file, and must return a working revision~\citep{cassano2024editevaluatingabilitylarge}. The edit interface determines how that revision is expressed. For example, search-and-replace tools let a model identify an old substring and supply its replacement, leaving the runtime to preserve the rest of the file~\citep{anthropic2026texteditor}.

Masked diffusion language models (MDLMs) offer another way to organize this work. Their bidirectional denoising can condition on both sides of an edit, keep observed source fixed, and fill several disjoint regions jointly~\citep{sahoo2024mdlm,nie2025largelanguagediffusionmodels,gong2025diffucoderunderstandingimprovingmasked}. These properties suggest that an editor could generate new code directly where it belongs. Yet a code-editing request supplies no masks: the model must first decide which parts of a valid program to expose for generation. Known-gap infilling methods address content and length once the sites are given~\citep{bavarian2022fim,wu2026dreamondiffusionlanguagemodels,kim2025anyorderflexiblelengthmasked,xu2026predictdontiterateefficient}; instructional editing adds the problem of allocating those sites.

This creates a tradeoff between preserving source and committing to structure. Whole-file rewriting leaves edit locations implicit but regenerates unchanged code. Search-and-replace avoids that regeneration by predicting exact source anchors. Locate-then-infill exposes editable regions before jointly generating their contents, while token-level editing makes analogous decisions at individual token positions. The more source the runtime preserves, the more consequential the model's explicit choices of edit count, sites, and boundaries become. Which interfaces allow diffusion models to use their infilling capability effectively?

\begin{figure*}[t]
\centering
\includegraphics[width=\linewidth]{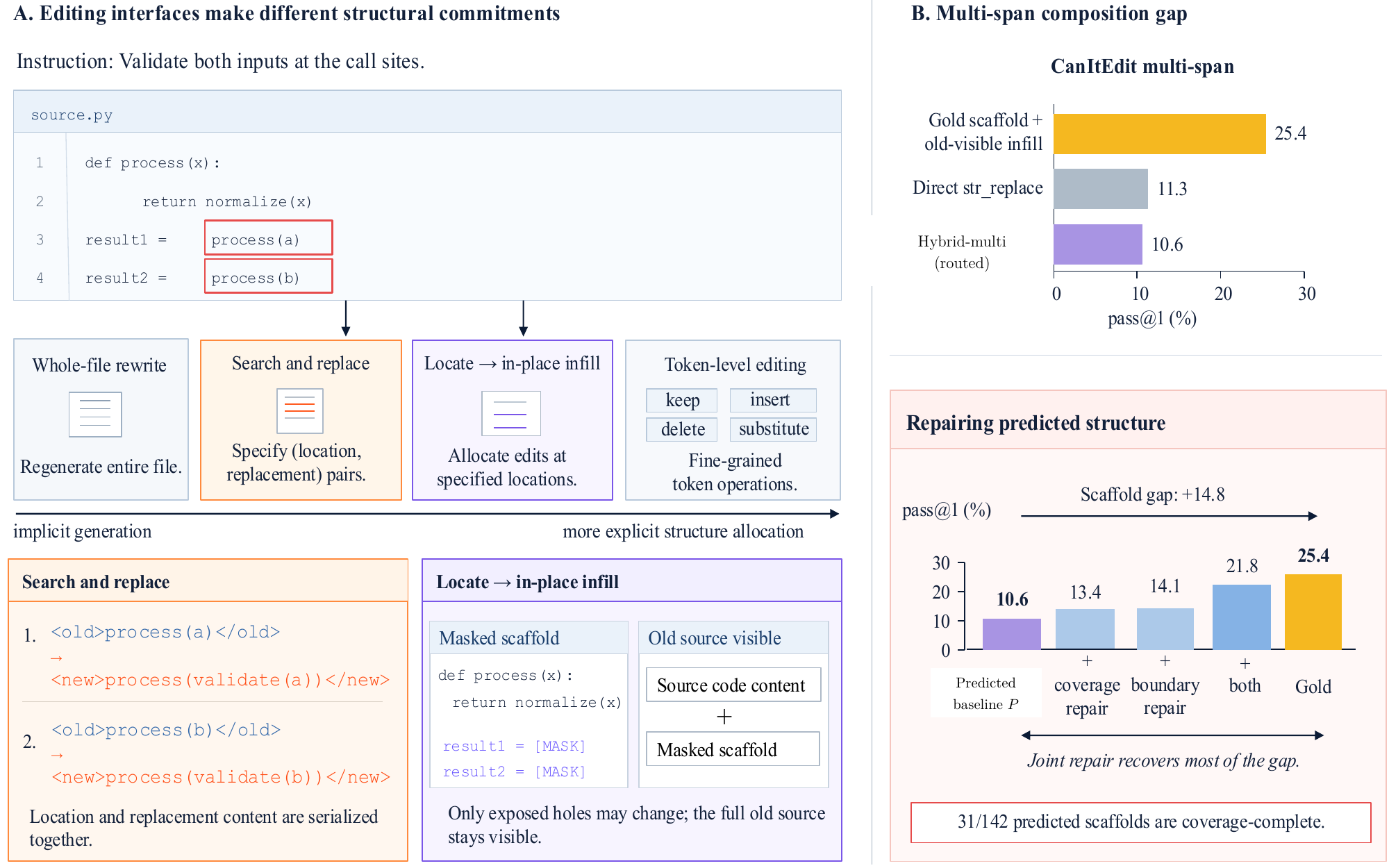}
\caption{Editing interfaces and the multi-span composition gap. (A) Four interfaces assign different structural decisions to the model and preservation to the runtime. (B) On 142 multi-span CanItEdit instances, gold-region old-visible infilling reaches 25.4\% pass@1, compared with 11.3\% for direct \strreplace{} and 10.6\% when predicted regions are composed with the infiller (Hybrid-multi). Matched oracle repairs of a separate predicted-hole baseline (P) expose complementary coverage and boundary errors. These repairs are diagnostics, not autonomous editors; routing and repair details appear in Appendix~\ref{app:composition}.}
\label{fig:main-method}
\end{figure*}

We investigate this question through implementations of all four interfaces, oracle-location controls, and controlled scaffold perturbations on CanItEdit. The central finding is a composition gap: an infiller can coordinate multiple edits when given the correct regions, yet lose this advantage when those regions must be predicted. Our diagnostics identify two distinct requirements. Missing regions prevent required changes, while overly broad holes increase the amount of correct code that must be reconstructed. Recovering coverage by widening holes can therefore undermine the capability it was intended to expose. Figure~\ref{fig:main-method} summarizes the interfaces and the evidence for this gap.

\paragraph{Contributions.}
\begin{itemize}[leftmargin=*]
  \item We formulate and evaluate four diffusion code-editing interfaces in terms of the structural decisions they assign to the model and the preservation guarantees they assign to the runtime.
  \item We establish a gap between known-site infilling capability and autonomous editing, using source-conditioning controls and composition with a predicted locator.
  \item We isolate coverage and boundary precision as separate requirements through matched scaffold interventions, providing a diagnostic framework for evaluating future diffusion editors.
\end{itemize}

\section{Related Work}
Masked diffusion language models denoise masked tokens with bidirectional context~\citep{sahoo2024mdlm,shi2024simplified,nie2025largelanguagediffusionmodels,ye2025dream7bdiffusionlarge}; DiffuCoder adapts this objective to code and initializes our fine-tunes~\citep{gong2025diffucoderunderstandingimprovingmasked}. Known-gap methods address infilling once sites are supplied: FIM fixes a contiguous gap~\citep{bavarian2022fim}, while DreamOn, FlexMDM, and PILL adapt its length~\citep{wu2026dreamondiffusionlanguagemodels,kim2025anyorderflexiblelengthmasked,xu2026predictdontiterateefficient}. Instructional editing instead receives an intact file and must infer which regions to change, possibly at several disjoint sites. EditPackFT and CanItEdit provide our training and evaluation setting~\citep{cassano2024editevaluatingabilitylarge}.

Edit-based models such as the Levenshtein Transformer, DiffusER, and Edit Flows make insertion, deletion, or substitution part of generation~\citep{gu2019levenshtein,reid2022diffuserdiscretediffusioneditbased,havasi2025editflowsflowmatching}. LLaDA2.2 extends token editing to Levenshtein operations within diffusion blocks to correct generated drafts and improve agentic reliability~\citep{bie2026llada22}. In a small trace census, most observed LLaDA2.2 writes fill masks; revisions of existing canvas tokens are less common and vary by task (Table~\ref{tab:llada22_ops}). Our token-level interface instead selects instruction-conditioned edits throughout a supplied program (Section~\ref{sec:interfaces}). More broadly, we study how the edit interface allocates structure and how that allocation composes with masked-diffusion generation. Appendix~\ref{app:related_expanded} expands these distinctions.

\section{Editing Interfaces}
\label{sec:interfaces}
An \emph{edit interface} is the contract between a model and a deterministic runtime. Given an instruction $q$ and source file $x$, the model emits $o$, and the runtime produces
\begin{equation}
  \hat y = R(x,o).
  \label{eq:runtime}
\end{equation}
Interfaces differ in the structure that $R$ consumes, including locations, boundaries, edit count, and replacement lengths. In the sparse interfaces studied here, the runtime fixes which source regions generation may change. An omitted region remains unchanged even if the generated content elsewhere is correct.

We evaluate the reconstructed file $\hat y$, rather than exact agreement between $o$ and a reference serialization. Different replacements or plans may implement the same instruction, while a fluent intermediate object may fail to apply. The interface therefore determines both the model's prediction problem and its failure surface: rewrite concentrates all decisions in generated code, whereas sparse interfaces delegate preservation to $R$ and expose progressively more structure for exact execution.

\paragraph{Rewrite.} The model emits the complete edited file, $o=\hat y$, and the runtime is the identity. Localization and preservation remain implicit in generation. Whole-file output is used in the CanItEdit evaluation protocol~\citep{cassano2024editevaluatingabilitylarge}.

\paragraph{Search-and-replace.} Also called serialized replacement in Figure~\ref{fig:main-method}, this interface emits old/new substring pairs $((s_1,t_1),\ldots,(s_k,t_k))$. Each pair is serialized inside \texttt{<old>} and \texttt{<new>} fields. The \strreplace{} runtime substitutes each uniquely matching $s_i$ with $t_i$ and fails on missing, ambiguous, or overlapping matches; it performs neither whitespace normalization nor fuzzy matching. Untouched source is preserved, while anchors, boundaries, edit count, and serialization become explicit commitments. Replacement length remains implicit in the generated $t_i$. This is the same basic old-string/new-string contract used by text-editor tools~\citep{anthropic2026texteditor}; our tags define the serialization used in these experiments.

\paragraph{Locate-then-infill.} This interface separates structure allocation from content generation. Stage~1 reads $(q,x)$ with numbered source lines and emits an ordered list of runnable headers: \texttt{replace lines $a$--$b$}, \texttt{delete lines $a$--$b$}, or \texttt{insert new lines after line $c$}, with a start-of-file insertion variant. A header may include a rationale, but the runtime consumes only the operation and coordinates; plans are not allowed to paste replacement code. A deterministic parser converts the headers into a scaffold. Unnamed source lines remain observed, replacement and deletion ranges are removed from the editable output, and replacement or insertion sites receive masked holes. The runtime does not infer missing edits or widen predicted ranges.

\begin{figure}[t]
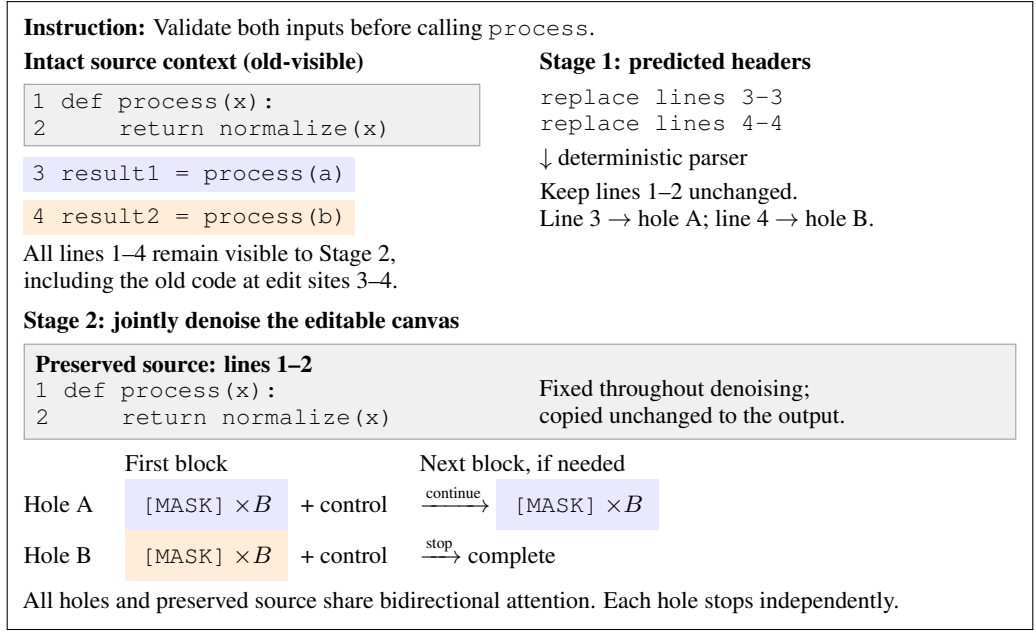

\centering
\setlength{\fboxsep}{6pt}
\fbox{\begin{minipage}{0.94\linewidth}
\small
\textbf{Instruction:} Validate both inputs before calling \texttt{process}.\\[3pt]
\begin{minipage}[t]{0.46\linewidth}
\textbf{Intact source context (old-visible)}
\\[3pt]
{\setlength{\fboxsep}{3pt}%
\fcolorbox{black!40}{black!6}{\begin{minipage}{\dimexpr\linewidth-2\fboxsep-2\fboxrule\relax}
\ttfamily\footnotesize
1 def process(x):\\
2 \phantom{xxxx}return normalize(x)
\end{minipage}}\\[3pt]
\colorbox{blue!9}{\texttt{\footnotesize 3 result1 = process(a)}}\\[2pt]
\colorbox{orange!15}{\texttt{\footnotesize 4 result2 = process(b)}}}\\[3pt]
{\footnotesize All lines 1--4 remain visible to Stage~2,\\
including the old code at edit sites 3--4.}
\end{minipage}\hfill
\begin{minipage}[t]{0.48\linewidth}
\textbf{Stage 1: predicted headers}\\[3pt]
\texttt{replace lines 3--3}\\
\texttt{replace lines 4--4}\\[3pt]
$\downarrow$ deterministic parser\\[3pt]
Keep lines 1--2 unchanged.\\
Line 3 $\to$ hole A; line 4 $\to$ hole B.
\end{minipage}
\par\medskip
\textbf{Stage 2: jointly denoise the editable canvas}\\[3pt]
{\setlength{\fboxsep}{4pt}%
\fcolorbox{black!40}{black!6}{\begin{minipage}{\dimexpr\linewidth-2\fboxsep-2\fboxrule\relax}
\textbf{Preserved source: lines 1--2}\\[2pt]
\begin{minipage}{0.49\linewidth}
\ttfamily\footnotesize
1 def process(x):\\
2 \phantom{xxxx}return normalize(x)
\end{minipage}\hfill
\begin{minipage}{0.48\linewidth}
\footnotesize
Fixed throughout denoising;\\
copied unchanged to the output.
\end{minipage}
\end{minipage}}}\\[4pt]
\begin{tabular}{@{}lll@{}}
 & First block & Next block, if needed \\
Hole A & \colorbox{blue!9}{\texttt{[MASK]} $\times B$}\; + control & $\xrightarrow{\text{continue}}$ \colorbox{blue!9}{\texttt{[MASK]} $\times B$} \\
Hole B & \colorbox{orange!15}{\texttt{[MASK]} $\times B$}\; + control & $\xrightarrow{\text{stop}}$ complete \\
\end{tabular}\\[3pt]
All holes and preserved source share bidirectional attention. Each hole stops independently.
\end{minipage}}
\caption{Worked locate-then-infill example with old-visible conditioning. Stage~2 sees all four original lines as read-only context, including the old code at replacement sites 3--4. In the separate editable canvas, gray lines 1--2 are preserved verbatim, while blue and orange lines 3--4 become holes A and B. Line numbers are annotations; headers abbreviate the serializer. Each hole starts with $B$ content masks and a control slot and can continue or stop independently; a final block can trim unused slots. All holes and fixed source tokens share bidirectional attention.}
\label{fig:infill_trace}
\end{figure}

Stage~2 receives the scaffold and jointly denoises its holes in one bidirectional canvas. Figure~\ref{fig:infill_trace} illustrates the process. An omitted site remains fixed source, so the filler cannot make a change there.

Training uses instruction, source, and target triples. For Stage~1 supervision, a deterministic line diff fixes each gold operation type and its coordinates; a teacher supplies only rationale text conditioned on the instruction, source, and target. At inference, the locator predicts headers from the instruction and numbered source alone. Applying these headers constructs the scaffold, and Stage~2 learns to fill its holes with target code. Intermediate mask-block states supervise generation and stopping at variable lengths. Appendix~\ref{app:interface_details} specifies the serialization and sampler.

The \emph{old-visible} variant supplies the intact source, including the text removed at replacement sites, as read-only context alongside the editable scaffold. Thus Stage~2 can condition on what each hole replaces without being allowed to alter observed tokens. Gold locations invoke the same parser and filler as a capability diagnostic; they oracleize edit count, sites, and boundaries, but not the generated content or its length. The autonomous composition experiment instead derives holes from predicted \strreplace{} \texttt{<old>} spans and discards their predicted replacements.

Each Stage~2 hole begins with a block of mask tokens. After denoising a block, a learned control slot stops that hole or appends another mask block; content tokens in the final block can terminate early and trim unused capacity. This block-decision mechanism allocates length independently for each predicted site while retaining full bidirectional attention across observed source and all holes. It is a decoding harness, not a block-causal attention pattern. Appendix~\ref{app:interface_details} gives the exact serializers, parser behavior, and decoding configuration.

\paragraph{Token-level editing.} We also explore an interface derived from Levenshtein alignment~\citep{gu2019levenshtein,reid2022diffuserdiscretediffusioneditbased}. Each source token is supervised to be kept, substituted, or deleted, while a predicted \expandtok{} operation allocates additional masked slots for insertions. Structure allocation therefore operates at individual token positions throughout the supplied file. Results appear in Section~\ref{sec:results}.

LLaDA2.2 also uses keep, substitute, delete, and insert operations, but applies them to intermediate generated drafts within diffusion blocks~\citep{bie2026llada22}. Its blockwise generation conditions on preceding clean blocks, and insertion/deletion is followed by padding or truncation to retain the block length. Our interface instead starts from an intact source program and predicts instruction-conditioned edits across the file, including sites far from its end. The distinction concerns both supervision and edit scope: reconstructing corrupted target tokens or refining a generated block does not directly specify which valid source tokens an instruction requires changing. Appendix~\ref{app:related_expanded} reports a small operation census with its tasks, decoder settings, and denominators.

\section{Experimental Setup}
\label{sec:setup}
We fine-tune DiffuCoder-Instruct~\citep{gong2025diffucoderunderstandingimprovingmasked} on the 20,602-example Python subset of EditPackFT. Evaluation uses 105 CanItEdit Python problems, each paired with descriptive and lazy instruction variants, for 210 instruction instances~\citep{cassano2024editevaluatingabilitylarge}. The primary metric is pass@1 on hidden tests. Gold line diffs partition these instances into 68 single-span and 142 multi-span edits; a replacement counts as one span, regardless of its internal line length.

We distinguish autonomous methods from oracle diagnostics. Rewrite, \strreplace{}, locate-then-infill, and token-level editing are autonomous DiffuCoder-family operating points, but their fine-tuning recipes are not fully matched; Appendix~\ref{app:extra_tables} records the line- and file-level recipe differences. Oracle scaffolds provide Stage~2 with gold edit count, locations, and boundaries while leaving content and variable length to the filler. Controlled scaffold perturbations then isolate sensitivity to missing regions, boundary errors, and false-positive holes while holding the filler fixed. A separate HumanEval-Infilling experiment supplies one contiguous gap and its oracle token length to compare middle and end placement; its \texttt{multi\_line} split is distinct from multi-span CanItEdit editing. Checkpoints, prompts, parsers, decode settings, and training details appear in Appendices~\ref{app:dataset_details}, \ref{app:decode}, and~\ref{app:interface_details}.

\section{Main Results}
\label{sec:results}

\subsection{Autonomous interfaces and oracle gaps}

Table~\ref{tab:interface-controlled} reports functional correctness and syntax failures for all four interfaces. Under the available DiffuCoder-family recipes, search-and-replace performs best overall and on multi-span edits. Its low syntax-error rate is consistent with preserving most source through exact replacements. However, successful reconstruction also depends on generating applicable anchors: parse and apply rates are 89\% and 75\%, respectively. These failures are distinct from Python syntax errors.

\begin{table}[t]
\centering
\small
\setlength{\tabcolsep}{3.4pt}
\caption{CanItEdit results (\%).
Overall, Single, and Multi report pass@1 ($\uparrow$); SynErr and Exc.\ are the percentages of evaluations whose harness status is \texttt{SyntaxError} and \texttt{Exception} ($\downarrow$).
Exception includes assertion failures and other runtime errors, and excludes SyntaxError and Timeout, so the columns do not sum to 100.
Autonomous DiffuCoder-family recipes differ, so their scores are operating points rather than a causal interface ranking.
Oracle rows supply edit positions; matched empty and old-visible fillers differ in source conditioning.
Dashes indicate unavailable split or exception results.}
\label{tab:interface-controlled}
\begin{tabular}{lrrrrr}
  \toprule
  \textbf{Interface / structure} & \textbf{Overall}$\uparrow$ & \textbf{Single}$\uparrow$ & \textbf{Multi}$\uparrow$ & \textbf{SynErr}$\downarrow$ & \textbf{Exc.}$\downarrow$ \\
  \midrule
  \multicolumn{6}{l}{\emph{Autonomous DiffuCoder-family editors}} \\
  Search-and-replace (\strreplace{}) & \textbf{23.3} & \textbf{48.5} & \textbf{11.3} & 5.7 & 68.6 \\
  Locate-then-infill & 12.9 & 27.9 & 5.6 & 15.2 & 70.0 \\
  Rewrite & 11.4 & 27.9 & 3.5 & 23.8 & 63.3 \\
  Token-level editing & 5.2 & --- & --- & 50.0 & --- \\
  \midrule
  \multicolumn{6}{l}{\emph{Structure and conditioning diagnostics}} \\
  Gold scaffold $+$ old-visible infill & 27.1 & 30.9 & \textbf{25.4} & 27.1 & 45.7 \\
  Gold scaffold $+$ matched empty infill & 20.5 & 33.8 & 14.1 & 26.7 & 52.9 \\
  Predicted \texttt{<old>} hybrid (Hybrid-multi) & 22.9 & 48.5 & 10.6 & 9.0 & 65.2 \\
  Token-level editing $+$ oracle positions & 30.0 & --- & --- & 37.6 & --- \\
  \bottomrule
\end{tabular}
\end{table}

\paragraph{Known-site capability does not transfer through the locator.}
With gold regions, old-visible infilling reaches 25.4\% multi-span pass@1, compared with 11.3\% for direct search-and-replace. This contrast establishes useful known-site capability; the gold scaffold supplies structural information unavailable to the separately trained autonomous editor. Composing the same filler with predicted regions in Hybrid-multi yields 10.6\%. The hybrid sends uniquely matched predicted multi-edit \texttt{<old>} regions to the filler and discards their predicted replacements. Only 53/142 multi-span examples enter that branch; predicted single edits and failed or non-unique matches retain the direct \strreplace{} reconstruction. This is an end-to-end routing experiment, with details in Appendix~\ref{app:composition}.

\paragraph{Token-level edits expose a similar localization gap.}
Autonomous token-level editing reaches 5.2\% pass@1, rising to 30.0\% with oracle edit positions. Syntax errors remain frequent even with those positions supplied (37.6\%). Finer edit granularity therefore exposes both a localization problem and a generation problem. The oracle token-level and line-level rows use different training and decoding procedures; their scores characterize each interface's conditional capability.

\subsection{Simple alternatives retain the gap}

\paragraph{Repeated replacement does not close the multi-span gap.}
A first-valid executor applies one uniquely matching edit per round and replans on the updated file. After five rounds, multi-span pass@1 remains 11.3\%, while single-span performance falls from 48.5\% for direct one-shot replacement to 36.8\% (Appendix~\ref{app:iterative}).

\paragraph{Matched Stage~1 backbone check.}
We train DiffuCoder, iLLaDA-8B-Instruct, and LLaDA-8B-Instruct as Stage~1 locators on EditPackFT plans and pass their predicted CanItEdit scaffolds to the same aligned old-visible DiffuCoder Stage~2 filler at step~1610. With the old source before the plan and a 160-token first plan block, the three pairings reach 13.3\% (28/210), 13.8\% (29/210), and 11.9\% (25/210) pass@1, respectively. Their multi-span scores are 12.0\%, 11.3\%, and 10.6\%. The matched DiffuCoder locator parses 91.0\% of plans but achieves 2.4\% exact line-operation agreement and predicts 1.89 edits on average against 2.78 gold edits. Under this matched layout, the three backbone pairings have similar functional scores and all retain weak localization (Appendix~\ref{app:external_locators}).

The next section isolates the structural requirements of the infiller by holding its checkpoint and decoder fixed while changing the scaffold.

\section{Why Does the Oracle Advantage Disappear?}
\label{sec:diagnostics}

\subsection{Known-site infilling is a useful primitive}
\label{sec:fim_placement}
\label{sec:oracle_infill}

\begin{figure}[t]
\centering
\includegraphics[width=0.78\linewidth]{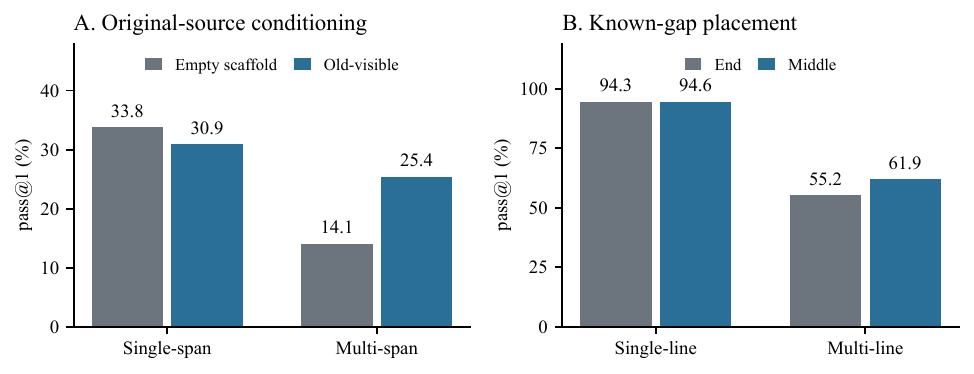}
\caption{Known-site generation controls (pass@1, \%). Left: original-source context helps the matched filler on multi-span CanItEdit. Right: after matched SFT, middle placement helps on harder, single-gap HumanEval-Infilling tasks with oracle gap length.}
\label{fig:infill_controls}
\end{figure}

Infilling can coordinate multiple edits when their sites are supplied, and retaining the original source strengthens that capability. Figure~\ref{fig:infill_controls} separates two questions: whether the filler benefits from seeing the removed source, and whether placing a known gap in the middle helps generation.

\paragraph{Original source helps coordinate multiple holes.}
Under matched Stage-2 training, supplying the intact source as read-only context raises multi-span pass@1 from 14.1\% to 25.4\%. Single-span scores are 33.8\% without this context and 30.9\% with it, so the gain is concentrated in multi-span editing. The old-visible filler can recover identifiers, expressions, and control flow removed from its editable canvas while keeping observed code fixed. Both variants receive gold locations and use the same variable-length block-decision procedure. Source visibility changes conditioning; the scaffold still determines which parts of the output may change.

\paragraph{Placement helps on harder known gaps after matched training.}
On HumanEval-Infilling, matched SFT yields 94.6\% for middle placement versus 94.3\% for end placement on \texttt{single\_line}, and 61.9\% versus 55.2\% on \texttt{multi\_line}. The much larger off-the-shelf single-line gap (93.6\% versus 17.7\%) is confounded by prompt and interface familiarity. Here both layouts receive the target length, and \texttt{multi\_line} denotes one contiguous gap. The matched result supports a placement benefit on the harder split, while the CanItEdit controls establish useful infilling with multiple holes and learned output lengths. The remaining question is whether a locator supplies regions that allow this capability to be used.

\subsection{Structure allocation is the bottleneck}
\label{sec:str_replace_audit}
\label{sec:loc_quality}

The predicted structures are usually incomplete. On multi-span examples, \strreplace{} \texttt{<old>} spans achieve 42.7\% mean gold-span overlap recall, and overlap every gold region in only 21.8\% of cases (31/142). Meanwhile, 45/142 predictions contain just one edit. Mean recall measures the fraction of gold regions overlapped per example; all-span coverage requires every region to be overlapped. Overlap alone does not ensure that a hole exposes every token requiring a change, but omitting a region entirely prevents the filler from editing it.

The Stage-1 planner exhibits the same structural deficit through a different representation. Across the evaluation set, 95.7\% of its plans parse, yet exact line-operation agreement is 13.3\%, and it predicts 1.55 edits on average against 2.78 gold edits. Thus well-formed headers frequently omit edits or choose incorrect ranges. Exact operation agreement and overlap recall measure different properties: the former checks operation and coordinates, while the latter checks intersection with gold regions. Together, these observations motivate separate tests of missing sites and inaccurate boundaries. Detailed locator and conditioning controls appear in Appendix~\ref{app:diag_tables}.

Matched repairs in Figure~\ref{fig:main-method} show that these errors are complementary. On the separate predicted-hole baseline $P$, multi-span pass@1 rises from 10.6\% to 13.4\% with coverage repair alone, 14.1\% with boundary repair alone, and 21.8\% with both, approaching the 25.4\% gold-scaffold result. Neither repair alone closes the composition gap.

\subsection{Coverage is necessary but not sufficient}
\label{sec:boundaries}
\label{sec:scaffold_corruption}
\label{sec:composition}
\label{sec:covered_subset}

\begin{figure}[t]
\centering
\includegraphics[width=0.98\linewidth]{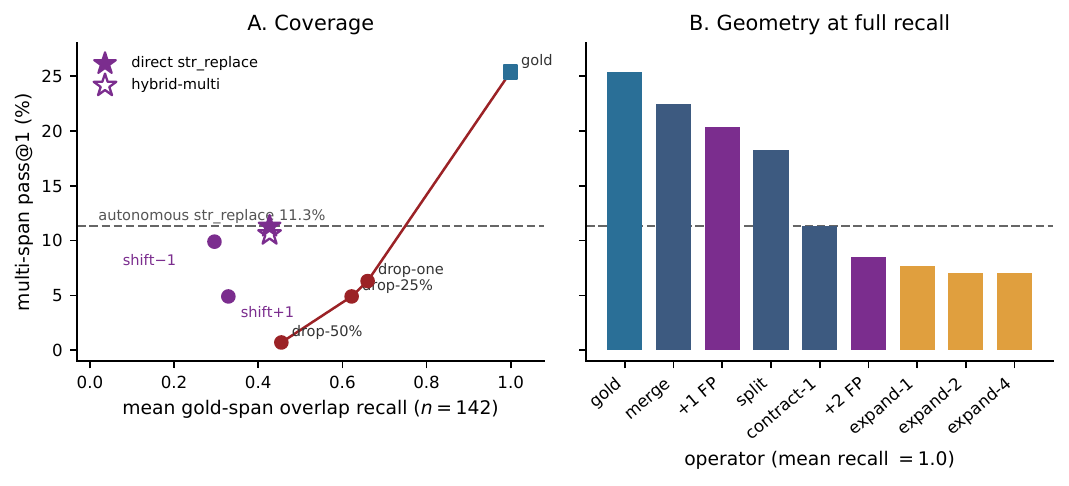}
\caption{Multi-span pass@1 of the fixed old-visible filler under controlled scaffold corruptions. Complete coverage alone is insufficient: expanding boundaries by one line performs worse than direct \strreplace{} (dashed line), whereas a single tight extra hole retains much of the gold-scaffold advantage.}
\label{fig:locator_quality}
\end{figure}

Matched scaffold corruptions separate missing regions from poor boundaries while holding the old-visible filler, prompt, and decoder fixed. Table~\ref{tab:scaffold_corruption} and Figure~\ref{fig:locator_quality} report multi-span pass@1. Dropping one required span is catastrophic (6.3\%), but making every hole one line wider also falls to 7.7\% despite retaining 100\% gold-span coverage. In contrast, one tight false-positive hole is comparatively tolerable (20.4\%); two reduce pass@1 to 8.5\%. Localization errors are asymmetric: missing a required region is highly damaging, but guaranteeing coverage with broad holes can erase the oracle advantage as well.

The interventions modify only scaffold geometry. \emph{Drop one} removes one required mask, leaving that old source region unchanged in the output. \emph{Expand $\pm1$} absorbs one kept source line on each available side of every gold range, requiring the filler to regenerate additional correct context in the editable canvas even though the original remains visible read-only. The false-positive conditions insert one or two tight masks over kept lines while preserving all gold masks. Gold-span recall therefore remains 100\% for both expansion and false positives, even though their generation burden differs. This design tests why widening every hole until all targets are covered is not an adequate locator policy.

\begin{table}[t]
\centering
\small
\setlength{\tabcolsep}{5pt}
\caption{Matched scaffold interventions on the old-visible filler. Recall is mean gold-span coverage on 142 multi-span CanItEdit examples; pass@1 is multi-span.}
\label{tab:scaffold_corruption}
\begin{tabular}{lrr}
  \toprule
  \textbf{Scaffold} & \textbf{Recall (\%)} & \textbf{pass@1 (\%)} \\
  \midrule
  Clean gold & 100 & \textbf{25.4} \\
  Drop one required span & 66.0 & 6.3 \\
  Expand each range $\pm 1$ & 100 & 7.7 \\
  Add one false-positive hole & 100 & 20.4 \\
  Add two false-positive holes & 100 & 8.5 \\
  \bottomrule
\end{tabular}
\end{table}

The fixed-filler interventions show that coverage and boundary precision impose distinct requirements. Missing holes prevent changes; oversized holes increase the correct context that must be regenerated. A single extra hole is less damaging under this protocol, but that tolerance decreases as extra holes accumulate. Natural locator outputs mix these errors, and Hybrid-multi's 10.6\% multi-span result shows that their composition does not recover the 25.4\% gold-scaffold capability. The Appendix provides the full corruption grid, complete-coverage subset, conditioning and length controls, sequential replacement results, and additional training interventions.

\subsection{Beyond code: targeted prose editing}
\label{sec:fineedit}
We construct sentence-level targeted edits from FineEdit Wiki examples~\citep{zeng2026bridgingeditinggapllms}, group near-duplicate examples before splitting, and fine-tune separate \strreplace{}, locator, and old-visible filler models from Dream-v0-Instruct-7B. We report a 64-example held-out probe using sentence-level pseudo-targets; Appendix~\ref{app:fineedit} details preprocessing, filtering, and evaluation.

\begin{table}[h]
\centering
\scriptsize
\setlength{\tabcolsep}{2.6pt}
\caption{FineEdit Wiki targeted editing on 64 grouped-split test examples. SARI, ROUGE-L, BLEU, and edit F1 use the $[0,1]$ scale; exact match, coverage, and preservation are percentages. Predicted and gold locations use the same old-visible filler and decoder. Coverage measures gold edits realized in the output; preservation measures source retained outside gold edit regions.}
\label{tab:fineedit_wiki}
\begin{tabular}{lrrrrrrr}
  \toprule
  \textbf{Interface / locations} & \textbf{Exact} & \textbf{SARI} & \textbf{ROUGE-L} & \textbf{BLEU} & \textbf{Edit F1} & \textbf{Coverage} & \textbf{Preserved} \\
  \midrule
  \strreplace{} & 0.0 & 0.411 & 0.882 & 0.795 & 0.061 & 6.8 & 97.5 \\
  Predicted locations $+$ infill & 3.1 & 0.437 & 0.807 & 0.728 & 0.376 & 45.3 & 82.3 \\
  Gold locations $+$ infill & \textbf{9.4} & \textbf{0.597} & 0.808 & 0.755 & \textbf{0.927} & \textbf{89.3} & 95.0 \\
  \bottomrule
\end{tabular}
\end{table}

Gold-location infilling reaches 0.927 edit F1 and 0.597 SARI, compared with 0.376 and 0.437 when the same filler uses predicted locations (Table~\ref{tab:fineedit_wiki}). Direct \strreplace{} applies on only 10.9\% of examples, so its high source preservation largely reflects unapplied edits.

These results reproduce the known-site versus predicted-site gap on sentence-level Wiki editing, though this 64-example slice cannot establish its size on the full test set. Exact match remains 9.4\% even with gold locations; SARI, edit F1, coverage, and preservation expose the editing behavior more directly here.

\section{Discussion: Preservation, Generation, and Cost}
\label{sec:discussion}

\paragraph{Preserved source is part of the interface's computational budget.}
Table~\ref{tab:interface_cost} compares what each interface generates to implement the same edit. Rewrite must reproduce unchanged code. Search-and-replace copies old anchors as well as generating their replacements. Locate-then-infill generates a plan and replacement content, while token-level editing predicts operations across the source and content at changed positions. Sparse interfaces can therefore reduce generated text when edits are small relative to the file. The corruption experiments expose the corresponding accuracy tradeoff: widening a hole spends more generation on previously correct code, and even full gold-span coverage can then yield poor functional accuracy.

\begin{table}[t]
\centering
\small
\caption{Generation responsibilities for a fixed edit. $U$ is unchanged output text, $V$ is replacement/insertion text, $A$ is the copied old-anchor text, and $H$ is the location-plan text, measured in tokens. Entries describe output content, excluding serialization/control overhead and repeated denoising passes; they are not runtime measurements.}
\label{tab:interface_cost}
\begin{tabular}{lll}
\toprule
\textbf{Interface} & \textbf{Text generated} & \textbf{Structure predicted} \\
\midrule
Rewrite & $U+V$ & Implicit in full output \\
Search-and-replace & $A+V$ & Exact source anchors \\
Locate-then-infill & $H+V$ & Regions and per-hole length \\
Token-level editing & Changed content & Operations across source tokens \\
\bottomrule
\end{tabular}
\end{table}

\paragraph{Output savings and decoding speed are different quantities.}
Fewer generated tokens can reduce the amount of content to denoise, but the cost also depends on canvas length, denoising rounds, and structural prediction. Locate-then-infill adds a planning pass, and old-visible conditioning adds source context. Although preserved source tokens stay fixed, their hidden states can still depend on changing holes under full bidirectional attention; source preservation alone does not guarantee exact KV-cache reuse. Our experiments measure functional accuracy. A speed comparison would require matched hardware, decoding budgets, context handling, and cache policies across interfaces.

\paragraph{Evaluate the structure--generation pair.}
An oracle-location score measures conditional infilling capability, while autonomous performance also depends on the regions exposed to the filler. Parseability, overlap recall, and pass@1 each reveal a different part of this interaction. The matched Stage~1 backbone check yields 11.9--13.8\% pass@1 with the same filler, alongside low exact line-operation agreement across all three locators. Reporting all-span coverage alongside boundary perturbations makes the composition gap inspectable: the locator must expose the required edits without imposing excessive reconstruction of unchanged code. This evaluation principle follows from the fixed-filler interventions and applies when assessing stronger locators or adaptive-length infillers.

\paragraph{Scope.}
\label{sec:limitations}
Our controlled study centers on Python file edits on CanItEdit, with a separate known-gap placement probe. A 64-example sentence-level FineEdit Wiki probe shows the same qualitative gold-versus-predicted location gap beyond code; its size on the full test set remains unknown. Repository-level and multi-file editing remain untested. Autonomous fine-tuning recipes differ. The controlled scaffold results concern our block-decision infiller. Other adaptive-length mechanisms may change its sensitivity to boundary errors~\citep{wu2026dreamondiffusionlanguagemodels,kim2025anyorderflexiblelengthmasked,xu2026predictdontiterateefficient}.

\section{Conclusion}
Diffusion code editing depends on how an interface divides structure allocation, generation, and source preservation. Across four interfaces, our study finds useful known-site infilling capability but a substantial gap when edit regions must be predicted. Matched scaffold interventions identify coverage and boundary precision as separate requirements: omitting required sites and widening otherwise complete holes can each erase the infilling advantage. A small sentence-level Wiki probe shows the same qualitative location gap beyond code. These controls provide a way to assess whether improvements in localization and generation translate into better autonomous editors.

\bibliography{references}
\bibliographystyle{iclr2027_conference}

\clearpage
\appendix
\section{Dataset Details}
\label{app:dataset_details}

This appendix provides detailed statistics for the EditPackFT Python training set
used in our experiments. All token-level measurements use the
model tokenizer; character-level and line-level measurements are computed on the raw
source strings.
Token-level keep/insert/delete/edit counts are a \emph{data description} obtained by
Levenshtein alignment of source and target (Appendix~\ref{app:levenshtein}); they do not
imply that the paper's headline methods train those token-level operations.

\begin{table}[h]
\centering
\caption{EditPackFT Python training set ($N{=}20{,}602$).
Token counts use the model tokenizer.}
\label{tab:dataset_summary}
\begin{tabular}{lrrr}
  \toprule
  \textbf{Statistic} & \textbf{Mean} & \textbf{Median} & \textbf{Max} \\
  \midrule
  Source tokens        & 174.6 & 174 & 546 \\
  Target tokens        & 204.1 & 204 & 730 \\
  Instruction tokens   &  28.0 &  27 & 118 \\
  Edit distance (tokens) & 49.5 &  26 & 722 \\
  Edit ratio ($\frac{\text{edit dist}}{\text{source len}}$) & 0.23 & 0.16 & 1.00 \\
  \bottomrule
\end{tabular}
\end{table}

\subsection{Sequence Length Distributions}

Table~\ref{tab:app_seq_len} reports the full percentile distribution of token-level
sequence lengths for source, target, and instruction fields.

\begin{table}[h]
\centering
\caption{Token-level sequence length distributions (EditPackFT Python, $N{=}20{,}602$).}
\label{tab:app_seq_len}
\resizebox{\linewidth}{!}{%
\begin{tabular}{lrrrrrrrrr}
  \toprule
  \textbf{Field} & \textbf{Mean} & \textbf{Std} & \textbf{Min} & \textbf{P25} & \textbf{Med} & \textbf{P75} & \textbf{P90} & \textbf{P95} & \textbf{Max} \\
  \midrule
  Source tokens    & 174.6 & 84.3 &   4 & 107 & 174 & 242 & 288 & 307 & 546 \\
  Target tokens    & 204.1 & 92.0 &   6 & 133 & 204 & 273 & 320 & 346 & 730 \\
  Instruction tokens & 28.0 &  3.9 &  23 &  25 &  27 &  29 &  32 &  35 & 118 \\
  Prompt + Source  & 202.6 & 84.5 &  28 & 135 & 202 & 270 & 316 & 336 & 577 \\
  Prompt + Target  & 232.1 & 92.2 &  32 & 161 & 232 & 301 & 348 & 374 & 755 \\
  \bottomrule
\end{tabular}%
}
\end{table}

Table~\ref{tab:app_char_line} reports character-level and line-level statistics for the
raw source and target code strings.

\begin{table}[h]
\centering
\caption{Character and line count distributions for source and target code.}
\label{tab:app_char_line}
\begin{tabular}{lrrrrrrr}
  \toprule
  \textbf{Field} & \textbf{Mean} & \textbf{Std} & \textbf{Min} & \textbf{Med} & \textbf{P75} & \textbf{P95} & \textbf{Max} \\
  \midrule
  Source chars   & 748.0 & 383.1 &  10 &  733 & 1033 & 1384 & 2937 \\
  Target chars   & 874.0 & 414.3 &  21 &  863 & 1175 & 1544 & 2947 \\
  Source lines   &  27.2 &  13.0 &   1 &   27 &   36 &   49 &   94 \\
  Target lines   &  31.1 &  14.0 &   1 &   31 &   41 &   55 &  164 \\
  Instruction chars & 44.9 & 17.7 &  16 &   41 &   51 &   74 &  444 \\
  \bottomrule
\end{tabular}
\end{table}

\subsection{Edit Operation Statistics}

Table~\ref{tab:app_edit_ops} reports the distribution of token-level edit operations obtained
from Levenshtein alignment of source and target token IDs.

\begin{table}[h]
\centering
\caption{Token-level edit operation distributions per example.}
\label{tab:app_edit_ops}
\begin{tabular}{lrrrrrrr}
  \toprule
  \textbf{Operation} & \textbf{Mean} & \textbf{Std} & \textbf{Min} & \textbf{Med} & \textbf{P75} & \textbf{P95} & \textbf{Max} \\
  \midrule
  \textsc{keep} tokens   & 161.5 & 81.7 & 1 & 160 & 227 & 293 & 379 \\
  \textsc{insert} tokens &  36.4 & 57.9 & 0 &  15 &  44 & 148 & 689 \\
  \textsc{delete} tokens &   6.9 & 22.6 & 0 &   0 &   4 &  35 & 525 \\
  \textsc{edit} tokens   &   6.3 & 14.6 & 0 &   2 &   6 &  28 & 279 \\
  Edit distance          &  49.5 & 63.1 & 1 &  26 &  61 & 175 & 722 \\
  Edit ratio             & 0.23  & 0.21 & 0.003 & 0.16 & 0.33 & 0.70 & 1.00 \\
  \bottomrule
\end{tabular}
\end{table}

\paragraph{Change type composition.}
Table~\ref{tab:app_change_types} shows the breakdown of examples by their edit operation
profile.

\begin{table}[h]
\centering
\caption{Distribution of examples by change type.}
\label{tab:app_change_types}
\begin{tabular}{lrr}
  \toprule
  \textbf{Change Type} & \textbf{Count} & \textbf{\%} \\
  \midrule
  Mixed (insert+delete+edit) & 14{,}362 & 69.7 \\
  Insert only                &  4{,}984 & 24.2 \\
  Delete only                &    993  &  4.8 \\
  Edit only (substitution)   &    263  &  1.3 \\
  \midrule
  Has insertions             & 17{,}426 & 84.6 \\
  Has deletions              &  7{,}603 & 36.9 \\
  Has substitutions          & 14{,}326 & 69.5 \\
  \bottomrule
\end{tabular}
\end{table}

\subsection{Insert Group Statistics}

\begin{table}[h]
\centering
\caption{Insert group statistics.  An insert group is a contiguous run of \textsc{insert} operations.}
\label{tab:app_insert_groups}
\begin{tabular}{lrrrrrrr}
  \toprule
  \textbf{Statistic} & \textbf{Mean} & \textbf{Std} & \textbf{Min} & \textbf{Med} & \textbf{P75} & \textbf{P95} & \textbf{Max} \\
  \midrule
  Groups per example & 2.3 & 2.5 & 0 &  2 &  3 &  7 &  31 \\
  Tokens per group   & 15.6 & 30.2 & 1 &  6 & 15 & 65 & 611 \\
  \bottomrule
\end{tabular}
\end{table}

The four-type Levenshtein training-instance accounting from earlier in-place experiments
(260{,}280 examples) is in Appendix~\ref{app:levenshtein}; it is not part of the Stage-1/2
or \strreplace{} recipes.

\section{Token-Level Levenshtein Families}
\label{app:levenshtein}

This section details the token-level interface: source and target
token IDs are aligned with Levenshtein dynamic programming, and the model is trained to
predict keep / substitute / insert / delete in place.
Only one vocabulary item is added, \expandtok{}, meaning ``insert one more \masktok{} slot
after me.''
Deletion reuses \texttt{EOS}: \texttt{EOS} on a source token deletes it; \texttt{EOS} on a
\masktok{} discards an extra slot (over-expand recovery).
Consecutive inserts form insert groups (Table~\ref{tab:app_insert_groups}).
The source program is treated as the diffusion state; loss is applied at response positions
rather than at randomly masked MDLM timesteps.

Table~\ref{tab:app_lev_families} compresses the nine training families.
Families 1--4 and 8--9 are precomputed; families 3 and 5--7 resample intermediate states
online.
We do not treat this bank as the training recipe for rewrite, \strreplace, or Stage-1/2.

\begin{table}[h]
\centering
\caption{Token-level Levenshtein training families (compressed). Families 1--4 were the
original four-type bank; 5--9 add partial-fill, recovery, residual, and no-op coverage.}
\label{tab:app_lev_families}
\small
\begin{tabular}{clp{0.62\linewidth}}
  \toprule
  \textbf{\#} & \textbf{Family} & \textbf{Supervised state} \\
  \midrule
  1 & Fill pair & Fully expanded \masktok{} slots; parallel fill; keep/edit/delete on source. \\
  2 & Expand trigger & Pre-mask source; \expandtok{} on the last source token before each insert group. \\
  3 & Intermediate expand & Random partition of each group; every slot expand vs.\ fill; $O(\log N)$ doubling. \\
  4 & Over-expand recovery & $N{+}1$ slots; extra slot labelled \texttt{EOS}. \\
  5 & Partial-fill & Some insert slots already committed; remaining masks filled. \\
  6 & Spurious-expand & Bogus masks on non-insert positions, labelled \texttt{EOS}. \\
  7 & Partial-commit residual & Some edit/delete already applied; realign residual, then another family. \\
  8 & Edit/delete-focus fill & Like fill, but loss only on edit and delete positions. \\
  9 & Full-keep & Source already equals target (no-op). \\
  \bottomrule
\end{tabular}
\end{table}

On the original four-type accounting, 20{,}602 EditPackFT Python pairs expand to 20{,}602 fill
pairs, 17{,}426 expand-trigger pairs (rows with insertions), 47{,}992 over-expand recovery
examples, and 174{,}260 lazily sampled intermediate examples (260{,}280 instances).
Rare operations are upweighted by inverse-frequency per-operation weights (keep/insert down,
edit/delete/expand up); we do not claim that weighting was material to CanItEdit pass@1.

\paragraph{Localization outcome.}
On CanItEdit, autonomous token-level editing reaches
5.2\% pass@1 with 50.0\% SyntaxError. Oracle edit positions raise pass@1 to 30.0\%,
while 37.6\% of outputs still have a SyntaxError. Thus localization accounts for a
large part of the gap, but known positions do not remove generation errors.
An earlier sister-repo exact-match diagnostic used a different protocol and found
about 30--32.5\% with oracle mask positions versus about 0\% with model-chosen
insert points; these are not the 210-instance pass@1 results.

\section{Decode Settings}
\label{app:decode}

Table~\ref{tab:app_decode} recaps the decode settings used for headline CanItEdit cells.
Stage-1 confidence $0.1$ and Stage-2 confidence $0.3$ are never mixed.
Block-decision / block-append is a length and stop harness on \emph{bidirectional}
masked-diffusion attention, not a block-causal attention ablation.
Single-span vs.\ multi-span splits use gold \texttt{line\_diff} insert spans (a replace is
one span).
\strreplace{} parse and apply failures are reported separately from SyntaxError.

\begin{table}[h]
\centering
\caption{Decode settings for headline CanItEdit cells.}
\label{tab:app_decode}
\small
\begin{tabular}{p{0.22\linewidth}p{0.20\linewidth}p{0.48\linewidth}}
  \toprule
  \textbf{Interface} & \textbf{Length unit} & \textbf{Other} \\
  \midrule
  Stage-1 plans
    & block-decision, size 32
    & confidence $0.1$; numbered old code in the best runs \\
  Stage-2 fill
    & block-decision, size 16
    & confidence $0.3$; joint denoising of all holes \\
  Rewrite SFT
    & fixed 256 or 512 steps
    & full-file generation; no Stage-1 plan \\
  \strreplace{} (headline)
    & block-append, step 1685
    & serialized $(s_i \to t_i)$; runtime apply-in-order \\
  \strreplace{} (budget)
    & 256-slot, step 1515
    & same interface, smaller generation budget \\
  \strreplace{} (pad)
    & random-pad, step 1685
    & same recipe as headline except padding \\
  Unified diff
    & DiffuCoder FT
    & weaker sibling; not centered in the main comparison \\
  \bottomrule
\end{tabular}
\end{table}

Table~\ref{tab:ckpt-identity} maps the role names used in Sections~\ref{sec:setup}--\ref{sec:results} to training steps and init paths.
Aligned empty-plan and old-visible Stage~2 fillers (Class~C and Table~\ref{tab:oracle_conditioning}) are separately trained from DiffuCoder-Instruct at \texttt{global\_step\_1610} (\texttt{empty\_plan} and \texttt{original\_source}); they are not the oracle Stage-2 specialist below.
The local $\pm 2$ and integrated-diff fillers are further aligned Stage-2 SFT runs from DiffuCoder-Instruct at the same step (\texttt{diff\_local\_old\_visible}, \texttt{integrated\_diff\_scaffold}).
Glue-fix and minimal-edit canonicalization reuse the integrated-diff step-1610 weights; they change only gold-scaffold construction (Section~\ref{sec:scaffold_construction}).
Qwen gold CanItEdit plans (\texttt{outputs/canitedit\_editing\_plans/plans.jsonl}) are
gold-location-stamped and are not an autonomous planner.

\begin{table}[h]
\centering
\small
\caption{Checkpoint identities for the roles named in the main text.
Decode protocol is in Table~\ref{tab:app_decode}; recipe confounds are in Table~\ref{tab:recipe-confounds}.}
\label{tab:ckpt-identity}
\begin{tabular}{p{0.24\linewidth}p{0.09\linewidth}p{0.58\linewidth}}
  \toprule
  \textbf{Role} & \textbf{Step} & \textbf{Init / path} \\
  \midrule
  Rewrite SFT
    & 640
    & DiffuCoder-Instruct; \path{ml-diffucoder/checkpoints/sft_editpackft_baseline_rewrite_8gpu} \\
  \strreplace{} SFT
    & 1685
    & DiffuCoder-Instruct; \path{checkpoints/str_replace_sft_block} (block-append) \\
  Shared Stage-1/2 multitask
    & 1935
    & oracle Stage-2 specialist 1610; \path{checkpoints/stage23_multitask_sft_from_stage3_step1610} \\
  Oracle Stage-2 specialist
    & 1610
    & DiffuCoder-Instruct; \path{checkpoints/stage3_oracle_scaffold_diffucoder_block_decision_ep5} \\
  \bottomrule
\end{tabular}
\end{table}

\section{Additional Failed-Intervention Details}
\label{app:failed}

Table~\ref{tab:what-failed} summarizes interventions that did not resolve structure allocation.

\begin{table}[h]
\centering
\small
\caption{Failed finer-grained interventions.
None became the method of the paper.}
\label{tab:what-failed}
\resizebox{\linewidth}{!}{%
\begin{tabular}{p{0.28\linewidth}p{0.34\linewidth}p{0.30\linewidth}}
  \toprule
  \textbf{Attempt} & \textbf{Result} & \textbf{Lesson} \\
  \midrule
  Token-level Levenshtein editing
    & Oracle masks $\sim$30--32.5\% exact match; model-chosen insert points $\sim$0\%.
    & Bottleneck is \emph{where} to insert, not fill given where. \\
  Header reweight, skeleton weight, header-only targets, multi-hunk oversampling
    & All failed location gates (exact line-op $>$25\%; mean edits toward gold 2.78). Header-only raised count but produced out-of-bounds locations.
    & Upweighting format tokens is not a substitute for learning the discrete structure those tokens encode. \\
  Frozen-backbone dense \texttt{ScaffoldHead}
    & CanItEdit exact line-op 0.48\%; in-domain exact event-tape 6.6\%. Stopped: no fill, no unfreeze.
    & Coarse metrics (nearby anchors, non-zero edits) can look calibrated while the scaffold remains unusable. \\
  OpenCoder 50/50 Stage-2 mix; hole markers; per-token DreamOn expand
    & Gold-loc / Qwen 38.1\% vs.\ ep5 39.5\%. Markers swapped EOS-error types; DreamOn SyntaxError $\sim$38\%.
    & More generic code did not raise the fill ceiling. Block-decision was the production length harness because it avoided the worst syntax collapse. \\
  Round-consistent / block-state-balanced Stage-2 sampler
    & Exact realized length $16.4\% \to 21.5\%$; under-generation $71.1\% \to 65.9\%$; pass@1 $27.1\% \to 23.8\%$; multi-span $25.4\% \to 19.0\%$. Long 1-hole ($33$--$64$ / $65+$) unchanged.
    & Matching gold trajectory states more closely improved length statistics, not functional editing (Table~\ref{tab:round_balanced}). \\
  \bottomrule
\end{tabular}%
}
\end{table}

\paragraph{Length-control variants.}
\label{app:length_control}
Hole markers around Stage-2 blocks did not improve pass@1; they flipped missing-\texttt{EOS}
errors into premature-\texttt{EOS}.
Force-masking the first gold \texttt{EOS} lowered pass@1 and raised SyntaxError.
Per-token DreamOn expand/delete was SyntaxError-heavy ($\sim$38\% in recorded locked-length
comparisons).
Under gold locations and privileged gold-aware Qwen plans, learned variable-length
block-decision already sits close to locked / oracle-length infill on the full CanItEdit
protocol (different Stage-2 weights; not mixed with the 25.4\% old-visible cell, and not
the same-checkpoint occupancy grid in Appendix~\ref{app:length_sensitivity}):
fixed-length locked DiffuCoder 46.7\% (SyntaxError 17.6\%);
block-append locked step 966 44.3\%;
the original block-decision joint dump 41.9\%;
production ep5 39.5\% and shared multitask 42.9\%.
The 2--7 point gap is much smaller than the drop from those oracle-location cells to
autonomous 12.9\%, so we do not treat hole-length allocation as the main explanation of
the locate-then-infill collapse.
A same-checkpoint old-visible forced-reference-occupancy grid is now in
Appendix~\ref{app:length_sensitivity}: $\delta{=}0$ is a non-monotonic intervention, not
an oracle-length upper bound.
On the glue-fixed integrated scaffold, forcing gold occupancy at inference with the same
step-1610 variable-length checkpoint scores 27.1\% overall (SyntaxError 40.0\%): the
124 unaffected evaluations drop from 34.7\% to 30.6\%, while the 86 EOF/no-newline
evaluations rise from 14.0\% to 22.1\%.
Canonicalized \emph{variable-length} decoding already obtains 28.1\%, so this mismatched
diagnostic is not a Stage-2 upper bound, and we do not infer from it that length control
is unimportant (Section~\ref{sec:scaffold_construction}).
Mixed block-size Stage-2 training is implemented but was not trained.

\paragraph{Trajectory-matched Stage-2 sampler.}
\label{app:round_balanced}
The aligned old-visible filler supervises one sampled gold block per hole, with independently sampled hole phases.
A round-consistent, block-state-balanced variant instead samples one global round per example, freezes completed holes, and reweights examples by $R{=}\max_h K_h$ so that later gold trajectory states are seen more often (same DiffuCoder-Instruct recipe, original-source context, 1610 optimizer steps).
Both still use gold committed prefixes.
Table~\ref{tab:round_balanced} is a gold-location diagnostic, not autonomous performance.
Exact-length and under-generation rates are among aligned completions ($152/210$ baseline; $135/210$ round-consistent).
Long 1-hole buckets are unchanged at $8.3\%$ ($33$--$64$ tokens) and $0\%$ ($65+$).
Teacher-forced continue accuracy is omitted from the main text: it moved only $57.3\% \to 58.8\%$.
We do not read this as a proof that Stage~2 cannot be compositionally robust, only that better matching gold trajectory states was insufficient under this recipe.

\begin{table}[h]
\centering
\small
\setlength{\tabcolsep}{3.6pt}
\caption{Round-consistent / block-state-balanced Stage-2 training versus the aligned old-visible baseline
(CanItEdit, gold locations, original-source context, unlocked block-decision).
Exact-length and under-generation rates are among aligned completions.
Rates are percentages.
Diagnostic, not autonomous performance.}
\label{tab:round_balanced}
\resizebox{\linewidth}{!}{%
\begin{tabular}{lrrrrrr}
  \toprule
  \textbf{Stage-2 training} & \textbf{pass@1} & \textbf{SynErr} & \textbf{1-span} & \textbf{Multi} & \textbf{Exact len} & \textbf{Under-gen} \\
  \midrule
  Aligned old-visible & 27.1 & 27.1 & 30.9 & 25.4 & 16.4 & 71.1 \\
  Round-consistent    & 23.8 & 29.0 & 33.8 & 19.0 & 21.5 & 65.9 \\
  \bottomrule
\end{tabular}%
}
\end{table}

\clearpage
\section{Old-visible Occupancy and Length Diagnostics}
\label{app:length_sensitivity}

These cells hold the aligned old-visible Stage~2 filler fixed
(\texttt{global\_step\_1610}, gold locations, no plan file, block size 16, 64 steps,
confidence 0.3, joint denoising).
\texttt{--lock-block-count} forces reference occupancy while the usual block-decision
machinery still generates content; a shared per-hole token offset $\delta$ is applied to
every hole in an example.
The unlocked cell is the existing 27.1\% dump, not a rerun.
This is a forced-reference-occupancy intervention, not an oracle-length upper bound
(Section~\ref{sec:oracle_infill}).
Jobs: length grid 137875; control-slot observer 137903.

\begin{table}[h]
\centering
\small
\setlength{\tabcolsep}{3.2pt}
\caption{Same-checkpoint forced-reference occupancy on the aligned old-visible filler
(CanItEdit).
Hole buckets are gold insert-span counts ($n{=}68/54/28/60$).
1-hole, multi, and per-count columns are pass@1 / SyntaxError.
Rates are percentages.
Diagnostic, not autonomous performance.}
\label{tab:oldvis_length_grid}
\resizebox{\linewidth}{!}{%
\begin{tabular}{lrrrrrrr}
  \toprule
  \textbf{Setting} & \textbf{pass@1} & \textbf{SynErr} & \textbf{1-hole} & \textbf{multi} & \textbf{2} & \textbf{3} & \textbf{$4+$} \\
  \midrule
  Unlocked & 27.1 & 27.1 & 30.9 / 35.3 & 25.4 / 23.2 & 44.4 / 11.1 & 7.1 / 28.6 & 16.7 / 31.7 \\
  $\delta{=}0$ forced occupancy & 28.1 & 42.4 & 47.1 / 32.4 & 19.0 / 47.2 & 33.3 / 35.2 & 21.4 / 60.7 & 5.0 / 51.7 \\
  $\delta{=}+1$ & 27.6 & 46.2 & 42.6 / 29.4 & 20.4 / 54.2 & 40.7 / 37.0 & 17.9 / 60.7 & 3.3 / 66.7 \\
  $\delta{=}+2$ & 24.8 & 50.5 & 41.2 / 38.2 & 16.9 / 56.3 & 35.2 / 38.9 & 14.3 / 60.7 & 1.7 / 70.0 \\
  $\delta{=}+4$ & 14.8 & 56.7 & 26.5 / 44.1 & 9.2 / 62.7 & 20.4 / 48.1 & 7.1 / 60.7 & 0.0 / 76.7 \\
  $\delta{=}-1$ & 4.3 & 71.0 & 1.5 / 69.1 & 5.6 / 71.8 & 11.1 / 70.4 & 3.6 / 67.9 & 1.7 / 75.0 \\
  $\delta{=}-2$ & 1.0 & 75.2 & 1.5 / 70.6 & 0.7 / 77.5 & 1.9 / 70.4 & 0.0 / 78.6 & 0.0 / 83.3 \\
  $\delta{=}-4$ & 2.4 & 75.2 & 2.9 / 67.6 & 2.1 / 78.9 & 3.7 / 61.1 & 0.0 / 82.1 & 1.7 / 93.3 \\
  \bottomrule
\end{tabular}%
}
\end{table}

\begin{table}[h]
\centering
\small
\caption{Paired unlocked $\to$ $\delta{=}0$ outcome churn on the same evaluation instances.
Pass is harness OK; SynErr is harness \texttt{SyntaxError} (IndentationError remains Other).
$99/210$ examples change outcome.}
\label{tab:oldvis_churn}
\begin{tabular}{lrrr}
  \toprule
  unlocked $\downarrow$ / $\delta{=}0$ $\rightarrow$ & Pass & SynErr & Other \\
  \midrule
  Pass   & 38 & 15 & 4 \\
  SynErr & 12 & 30 & 15 \\
  Other  &  9 & 44 & 43 \\
  \bottomrule
\end{tabular}
\end{table}

\begin{table}[h]
\centering
\small
\caption{Unlocked $\to$ $\delta{=}0$ churn by gold hole count.
The 1-hole gain is a SyntaxError rescue; 2-hole and $4+$ examples donate successful
unlocked trajectories into SyntaxError.}
\label{tab:oldvis_churn_holes}
\begin{tabular}{lrrrr}
  \toprule
  \textbf{Holes} & $n$ & changed & Pass$\to$SynErr & SynErr$\to$Pass \\
  \midrule
  1  & 68 & 28 & 0 & 9 \\
  2  & 54 & 21 & 9 & 3 \\
  3  & 28 & 17 & 0 & 0 \\
  $4+$ & 60 & 33 & 6 & 0 \\
  \bottomrule
\end{tabular}
\end{table}

\begin{table}[h]
\centering
\small
\caption{Unlocked old-visible pass@1 and SyntaxError by gold hole length.
The $65+$ 1-hole collapse ($0\%$ / $70\%$) is unchanged under $\delta{=}0$.
Bucket boundaries are empirical for this checkpoint/protocol, not a universal threshold.}
\label{tab:oldvis_size}
\begin{tabular}{llrrr}
  \toprule
  \textbf{Slice} & \textbf{Max gold hole} & $n$ & \textbf{pass@1} & \textbf{SynErr} \\
  \midrule
  1-hole & 1--16  & 14 & 57.1 & 14.3 \\
  1-hole & 17--32 & 22 & 54.5 & 18.2 \\
  1-hole & 33--64 & 12 & 8.3  & 33.3 \\
  1-hole & $65+$  & 20 & 0.0  & 70.0 \\
  \midrule
  All examples & $65+$ & 74 & 6.8 & 47.3 \\
  Total gold tokens $129+$ & --- & 60 & 8.3 & 45.0 \\
  1-hole, total gold tokens $129+$ & --- & 10 & 0.0 & 80.0 \\
  \bottomrule
\end{tabular}
\end{table}

Natural realized length is a post-hoc diagnostic on unlocked completions (no re-decoding).
Alignment was available for $152/210$ examples by \texttt{src\_idx}.
Realized token lengths come from a Myers-style line diff against the completion; block
counts are occupancy-based (\texttt{oracle\_block\_sizes}, block size 16), not decoder
\texttt{n\_blocks}.
Aligned Pass examples ($n{=}51/57$) have 45.1\% exact token length, 72.5\% exact blocks,
and mean max $|\Delta|$ 5.1 tokens.
Aligned SyntaxError examples ($n{=}40/57$) have 0.0\% exact token length, 5.0\% exact
blocks, mean max $|\Delta|$ 79.9, and 100\% under-generation of at least one span.
Naturally exact token length: 25 examples $\to$ 23 Pass, 0 SyntaxError, 2 Other.
For malformed programs, inferred under-allocation may partly reflect a broken file whose
alignment is shorter; the cleanest evidence is the exact-length subset having 0 SyntaxErrors.

\begin{table}[h]
\centering
\small
\caption{Teacher-forced control-slot agreement on the $\delta{=}0$ trajectory
(observer-only: gold stop/continue still forced).
Continue accuracy is agreement on non-final reference blocks; stop accuracy is agreement
on the gold final block.
These probabilities are not unconditional length-prediction accuracy.}
\label{tab:oldvis_ctrl}
\begin{tabular}{lrrrrr}
  \toprule
  \textbf{Split} & $n$ & cont.\ acc. & stop acc. & all holes & mean $P(\text{cont.})$ \\
  \midrule
  Overall & 210 & 57.3 & 73.3 & 50 & 0.281 \\
  1-hole & 68 & 32.4 & 76.5 & 25 & 0.247 \\
  2-hole & 54 & 69.9 & 71.3 & 16 & 0.308 \\
  3-hole & 28 & 51.5 & 70.2 & 5 & 0.312 \\
  $4+$ & 60 & 65.0 & 74.0 & 4 & 0.270 \\
  Multi SynErr & 67 & 74.0 & 63.3 & 5 & 0.375 \\
  Span $65+$ & 74 & 50.4 & 74.1 & 4 & 0.268 \\
  \bottomrule
\end{tabular}
\end{table}

Continue accuracy is $57.3\%$ overall (481/840) and only $32.4\%$ (55/170) on 1-hole
examples; stop accuracy is $73.3\%$ (419/572).
Only $50/210$ examples have every hole's forced stop/continue decision matching the
model's preferred class.
Mean $P(\text{continue})$ on the gold final block is 0.281 overall.
The dominant disagreement is early stop on non-final reference blocks, not a strong
desire to continue past the gold end.

\clearpage
\section{HumanEval-Infilling Placement Probe}
\label{app:humaneval_fim}
\label{app:fim_placement}

This appendix records a training-free placement probe on HumanEval-Infilling~\citep{chen2021codex,bavarian2022fim}, and a later matched-SFT comparison that holds the serialization fixed (Section~\ref{sec:fim_placement}).
It is \emph{not} a CanItEdit multi-span result, not a block-causal attention ablation, and not a test of variable-length control.
\texttt{multi\_line} is contiguous known-gap infilling, not multi-edit instructional editing.
All layouts below use a bidirectional DiffuCoder denoiser; they differ in where the gold-length mask run sits and, for the off-the-shelf end cell, in prompt format.

\paragraph{Protocol.}
Model: off-the-shelf \texttt{apple/DiffuCoder-7B-Instruct} for Table~\ref{tab:fim_placement}; matched SFT uses the same backbone after five epochs (Section~\ref{sec:fim_placement}).
Splits: HumanEval-Infilling \texttt{single\_line} ($n{=}1033$) and, for matched SFT, \texttt{multi\_line} ($n{=}5815$).
Decode: span denoiser, gold mask budget, expand/delete disabled, 64 steps, confidence $0.3$, one transfer token.
Completions are scored by concatenating $\texttt{prefix}+\texttt{completion}+\texttt{suffix}$ and running the reference unit tests (pass@1).
Exact-match is a looser diagnostic: it compares \texttt{completion.strip()} to the gold middle, which drops leading indentation and trailing newlines.

\paragraph{Layouts.}
\emph{Middle} is standard FIM: $\texttt{prefix}\,\masktok{}\times L\,\texttt{suffix}$, where $L$ is the gold token length of the missing span.
\emph{End-prompted} puts prefix and suffix in a chat user turn (fill in the missing code that belongs between PREFIX and SUFFIX; return only the missing code) and places the same gold-length mask run in the assistant canvas, so the hole is at the end of the sequence rather than in situ.
Suffix tokens remain visible in both layouts because attention is bidirectional; off the shelf, the contrast is in-place occupancy versus serialized generation, not access to the suffix.
That off-the-shelf contrast is not the matched-SFT comparison in Section~\ref{sec:fim_placement}.
An unprompted $\texttt{prefix}+\texttt{suffix}+\masktok{}$ layout and a FIM-sentinel layout are omitted from Table~\ref{tab:fim_placement}: they are not fair instruction-conditioned controls for this Instruct checkpoint.

\begin{table}[h]
\centering
\small
\caption{HumanEval-Infilling \texttt{single\_line} placement probe (DiffuCoder-Instruct, gold-length masks, $n{=}1033$).
The end-prompted row is the off-the-shelf protocol; it is not the matched-SFT end arm in Table~\ref{tab:fim_matched}.
pass@1 is the functional metric.
Rates are percentages.}
\label{tab:fim_placement}
\begin{tabular}{llrr}
  \toprule
  \textbf{Placement} & \textbf{Layout} & \textbf{exact-match} & \textbf{pass@1} \\
  \midrule
  Middle & $\texttt{prefix}\,\masktok{}\,\texttt{suffix}$ & 79.2 & 93.6 \\
  End-prompted & chat(PREFIX, SUFFIX) then $\masktok{}$ & 31.9 & 17.7 \\
  \bottomrule
\end{tabular}
\end{table}

\paragraph{Why exact-match and pass@1 disagree.}
On end-prompted, 175/1033 completions are strip-equal to gold yet fail tests: 156 miss a trailing newline (often gluing the next suffix line onto the generated line) and 19 differ only in indent.
Middle has zero such strip-equal failures; instead 149 programs pass tests with a different surface string than gold.
We therefore do not treat exact-match as a tighter metric than pass@1.

\paragraph{Off-the-shelf reading.}
Middle $\gg$ end-prompted on pass@1 for this Instruct checkpoint under the end-prompted protocol.
When the span is already known, putting the hole in its in-file position is strongly preferred \emph{off the shelf}.
That preference is not by itself evidence that locate-then-infill will succeed on CanItEdit, and it does not test joint multi-span infill.
Because the mask budget is gold, it also does not test a variable-length controller.
It does show that the CanItEdit bottleneck is not an inability to infill a known contiguous span.

\paragraph{Matched SFT.}
The matched arms initialize from the same DiffuCoder-Instruct checkpoint, train for five epochs (final step 315), and use paired EditPackFT single-span replacements with oracle target length.
Both arms serialize prefix/suffix with \texttt{[LEFT]}/\texttt{[RIGHT]} markers; the end arm places the gold-length mask run after a \texttt{[REPLACEMENT]} field rather than between prefix and suffix.
Table~\ref{tab:fim_matched} reports both HumanEval-Infilling splits; Table~\ref{tab:fim_matched_edit} reports held-out EditPackFT and the gold-location CanItEdit single-span diagnostic.
Matched SFT nearly ties \texttt{single\_line} ($94.6\%$ vs.\ $94.3\%$) but leaves a residual middle advantage on \texttt{multi\_line} ($61.9\%$ vs.\ $55.2\%$).
The matched-SFT CanItEdit $47.0\%$ cell is $n{=}66$ canonical single-span fills at oracle length; variable-length gold-scaffold Stage~2 evaluates the full evaluation set.

\begin{table}[h]
\centering
\small
\caption{HumanEval-Infilling with oracle span length. Off-the-shelf end uses the end-prompted chat protocol; the matched-SFT arms share one serialization. Rates are percentages.}
\label{tab:fim_matched}
\begin{tabular}{lrrr}
  \toprule
  \textbf{Evaluation} & \textbf{Middle} & \textbf{End} & \textbf{$\Delta$} \\
  \midrule
  Off-the-shelf \texttt{single\_line} pass@1 & 93.6 & 17.7 & $+75.9$ \\
  Matched-SFT \texttt{single\_line} pass@1 & 94.6 & 94.3 & $+0.3$ \\
  \quad exact match & 84.2 & 82.5 & $+1.7$ \\
  Matched-SFT \texttt{multi\_line} pass@1 & 61.9 & 55.2 & $+6.7$ \\
  \quad exact match & 37.3 & 33.1 & $+4.2$ \\
  \bottomrule
\end{tabular}
\end{table}

\begin{table}[h]
\centering
\small
\caption{Matched-SFT middle versus end on held-out EditPackFT ($n{=}400$) and gold-location CanItEdit canonical single-span fills ($n{=}66$), both with oracle span length.}
\label{tab:fim_matched_edit}
\begin{tabular}{llrr}
  \toprule
  \textbf{Evaluation} & \textbf{Metric} & \textbf{Middle} & \textbf{End} \\
  \midrule
  EditPackFT & exact match & 19.2 & 19.8 \\
   & syntax-ok & 97.0 & 96.8 \\
   & mean token distance & 12.2 & 12.2 \\
  CanItEdit single-span & pass@1 & 42.4 & 47.0 \\
   & string exact match & 24.2 & 27.3 \\
   & SyntaxError & 28.8 & 25.8 \\
  \bottomrule
\end{tabular}
\end{table}

\clearpage
\section{Expanded Related-Work Distinctions}
\label{app:related_expanded}

\paragraph{Masked diffusion language models.}
Masked diffusion language models replace tokens with an absorbing mask state and train a denoiser to reverse that corruption~\citep{sahoo2024mdlm,shi2024simplified}. LLaDA and Dream scale this objective to large language models~\citep{nie2025largelanguagediffusionmodels,ye2025dream7bdiffusionlarge}; iLLaDA updates the architecture and post-training recipe and supports variable-length continuation by appending mask blocks~\citep{nie2026improvedlargelanguagediffusionmodels}. DiffuCoder studies masked-diffusion post-training for code and initializes our fine-tunes~\citep{gong2025diffucoderunderstandingimprovingmasked}. DreamCoder is a separate open diffusion coder~\citep{xie2025dreamcoder7bopendiffusion}. Stable-DiffCoder~\citep{fan2026stablediffcoder} is an external rewrite reference from a different model and decoding stack.

\paragraph{Instructional code editing.}
Instructional editing supplies a request and an intact source file. Autoregressive code models~\citep{chen2021codex,lozhkov2024starcoder2} handle variable length through sequential decoding; EditPackFT and CanItEdit provide commit-derived training pairs and a Python evaluation protocol~\citep{cassano2024editevaluatingabilitylarge}. Whole-file rewrite, unified diffs, search-and-replace, and locate-then-infill expose different structural commitments to their runtimes. SWE-bench studies the related repository-level setting~\citep{jimenez2024swebench}, which is outside our file-level evaluation.

\paragraph{Edit-based generation.}
The Levenshtein Transformer, DiffusER, and Edit Flows generate through insertion, deletion, or substitution operations~\citep{gu2019levenshtein,reid2022diffuserdiscretediffusioneditbased,havasi2025editflowsflowmatching}. These approaches motivate structure as a generative primitive. LLaDA2.2 applies Levenshtein editing to intermediate drafts within diffusion blocks, with operation labels derived from LCS alignment and padding or truncation after edits to retain block length~\citep{bie2026llada22}. It targets error correction during generation and agentic reliability, whereas our token-level interface selects instruction-conditioned edits throughout a supplied source file. Our token-level Levenshtein variant instantiates this idea on a masked diffusion model; allocating edit and insertion positions is a major measured bottleneck (Appendix~\ref{app:levenshtein}).

\paragraph{LLaDA2.2 operation census.}
We inspected 20 greedy generation traces, five each for mathematics, prose, code writing, and code editing. Decoding used temperature~0, block size~32, confidence threshold~0.5, editing threshold~0.0, and a 128-token generation cap, matching the \texttt{infer\_22.py} settings. Mask-to-token (M2T) fills an empty canvas slot; replacement (T2T), deletion, and insertion revise the evolving canvas. Counts include repeated writes during generation and can therefore exceed final output length. Table~\ref{tab:llada22_ops} reports per-trace means.

\begin{table}[h]
\centering
\small
\setlength{\tabcolsep}{3.5pt}
\caption{LLaDA2.2 operation counts per greedy trace ($n{=}5$ per task). Structural share is replacement, deletion, and insertion divided by all writes.}
\label{tab:llada22_ops}
\begin{tabular}{lrrrrrr}
  \toprule
  \textbf{Task} & \textbf{Output tokens} & \textbf{M2T} & \textbf{Replace} & \textbf{Delete} & \textbf{Insert} & \textbf{Structural share} \\
  \midrule
  Mathematics & 121 & 137 & 4.4 & 1.6 & 1.2 & 5.0\% \\
  Prose & 107 & 124 & 9.0 & 2.8 & 1.6 & 9.7\% \\
  Code writing & 128 & 136 & 1.4 & 0.2 & 0.2 & 1.3\% \\
  Code editing & 104 & 117 & 3.4 & 1.2 & 0.4 & 4.1\% \\
  \bottomrule
\end{tabular}
\end{table}

Across the 20 traces, 2,569 of 2,706 writes (94.9\%) fill masks; replacement, deletion, and insertion account for 91, 29, and 17 writes. At least one structural operation occurs in 3/5 mathematics, 5/5 prose, 2/5 code-writing, and 4/5 code-editing traces. All five code-writing traces reach the 128-token cap, and three have no structural operation. The mathematics mean is influenced by one trace with 16 replacements, six deletions, and four insertions; two other mathematics traces have none. The code-editing prompts still generate from masks rather than locate changes in supplied source. This small census characterizes observed canvas revision under these prompts, not source-file localization or a stable population frequency.

\paragraph{Length harness versus block-causal attention.}
A different intervention from DreamOn sentinels changes the \emph{attention pattern}.
Block-causal attention (causal across blocks, bidirectional within a block) is an architectural constraint on information flow.
We do not train or evaluate a block-causal backbone in this paper.

What our experiments \emph{do} use for variable length is a length harness on top of full bidirectional \dllm{} attention.
\emph{Block-append} seeds a hole with a fixed-size block of masks, denoises that block, and concatenates another block if the hole is not yet finished.
\emph{Block-decision} additionally asks the model for an explicit stop/continue decision after each block.
In both cases the ``block'' is a generation and length-allocation unit---how many masks to place next---not a causal attention mask.
The denoiser still attends to the entire sequence.
Conflating this harness with block-causal attention would misstate both the decoder and the open question of whether full bidirectionality is necessary for editing.
Stable-DiffCoder~\citep{fan2026stablediffcoder} is a separate block-diffusion model family; we use it only as an off-the-shelf rewrite reference.

\paragraph{FIM versus instructional editing.}
Some FIM curricula permute prefix, suffix, and middle, and some training recipes include several holes, but the hole locations are still provided as markers rather than predicted from an editing instruction.
Adaptive-length methods---DreamOn expand/delete, FlexMDM insertion expectations, and PILL length probing~\citep{wu2026dreamondiffusionlanguagemodels,kim2025anyorderflexiblelengthmasked,xu2026predictdontiterateefficient}---then relax the remaining unknown, the span's token length.
DreamOn and the HumanEval-Infilling single-line / multi-line splits are contiguous known-gap infilling; multi-line is not multi-edit instructional editing.
FlexMDM can insert tokens at arbitrary positions in its generative formulation; its HumanEval-infilling evaluation still supplies the site via prefix/suffix separators.
PILL's appendix jointly infills several disjoint MBPP gaps, which is relevant to our Stage~2 setting, but those missing positions are still provided by construction.
A bidirectional \dllm{} can in principle denoise several disjoint holes jointly, which is the appeal of infilling as an edit interface.
That appeal assumes the holes are the right holes.
A HumanEval-Infilling \texttt{single\_line} placement probe (Appendix~\ref{app:humaneval_fim}) finds that off-the-shelf in-place middle infill substantially outperforms prompted end generation of the same gold-length span, even though suffix tokens remain visible in both layouts.
Matched SFT on a shared \texttt{[LEFT]}/\texttt{[RIGHT]}/\texttt{[REPLACEMENT]} formulation nearly eliminates that \texttt{single\_line} gap, but leaves a residual middle advantage on \texttt{multi\_line} (Section~\ref{sec:fim_placement}).
The probe does not test joint multi-span infill, and it does not test a variable-length controller.

\paragraph{Edit Flows versus source-conditioned editing.}
Edit Flows generate by moving through sequence space via insertions, deletions, and substitutions; they are not, in the form studied here, a comparison of rewrite versus patch versus locate-then-infill given an instruction and a source file.
Masked diffusion, conversely, iteratively refines a fixed set of slots unless a separate length mechanism (DreamOn sentinels, FlexMDM insertions, PILL-style length probes, or a block harness) is added.

\paragraph{Other distinctions.}
Stable-DiffCoder is not interchangeable with DiffuCoder-Instruct SFT cells (different family and decode stack).
SWE-bench~\citep{jimenez2024swebench} is a related repository-level protocol that we do not evaluate.
Token-level Levenshtein editing is a design point on a masked \dllm{}: choosing edit and insertion positions is a major bottleneck, and the oracle-position result still leaves substantial generation errors.

\section{Interface Implementation Details}
\label{app:interface_details}

\begin{table}[h]
\centering
\small
\caption{Structural commitments consumed explicitly by each runtime. A dash means that the decision remains latent in the generated output.}
\label{tab:structural-burden}
\setlength{\tabcolsep}{3.5pt}
\begin{tabular}{lcccccc}
\toprule
Interface & Explicit & Exact & Edit & Length & Multi-span & Source copied \\
 & location & boundary & count & allocation & serialization & by runtime \\
\midrule
Whole-file rewrite & --- & --- & --- & --- & --- & --- \\
Search-and-replace & \checkmark & \checkmark & \checkmark & --- & \checkmark & \checkmark \\
Locate-then-infill & \checkmark & \checkmark & \checkmark & \checkmark & \checkmark & \checkmark \\
Token-level in-place & \checkmark & \checkmark & \checkmark & \checkmark & \checkmark & \checkmark \\
\bottomrule
\end{tabular}
\end{table}

\paragraph{\strreplace{} serialization and parser.}
In our implementation, the model serializes replacements as a sequence of
\texttt{<edit><old>}$s_i$\texttt{</old><new>}$t_i$\texttt{</new></edit>}
blocks inside an \texttt{<edits>} wrapper.
The runtime $R_{\mathrm{sr}}$ locates each $s_i$ as an exact unique substring of the current file and substitutes $t_i$.
There is no whitespace normalization and no fuzzy match.
Length control applies to the serialized document, not to holes inside $x$: we train ordinary masked diffusion on one contiguous edit span, using either a fixed slot budget or the same block-append / block-decision harness as other generative interfaces.

\paragraph{Stage-1 headers.}
Gold training targets stamp gold hunk line numbers and ask a teacher only for rationale; plans must not paste new code.
At inference the source is presented with line numbers so that headers can refer to them.
Each edit is a parseable header of one of
\begin{center}
\texttt{Edit $N$: replace lines $a$--$b$ of the old file --} \textit{rationale}\\
\texttt{Edit $N$: delete lines $a$--$b$ of the old file --} \textit{rationale}\\
\texttt{Edit $N$: insert new lines after line $c$ --} \textit{rationale}
\end{center}
(with a start-of-file variant of the insert header).
The rationale is optional for the runtime; the headers are not.
A deterministic parser $\Pi$ turns $o_{\mathrm{plan}}$ and $x$ into a keep/delete/insert scaffold:
source lines not named by any header are kept; replace/delete ranges become deleted spans; insert and replace sites become holes of \masktok{} slots.
The runtime does not guess missing hunks or widen ranges for recall.

\paragraph{Integrated gold scaffolds.}
The aligned integrated-diff diagnostic serializes gold hunks into one canvas:
unchanged lines stay as ordinary source; replacements appear as read-only \texttt{-OLD} beside a \texttt{+[MASK]} hole; insertions are a hole with no minus line; deletions are minus lines with no hole.
Deterministic reconstruction drops minus lines and strips a leading \texttt{+}.
CanItEdit sources have no final newline; the original serializer could glue the last source line onto the following mask, so gold reconstruction failed on 43/105 programs.
Glue-fix newline-terminates minus records so that reconstruction round-trips.
Minimal-edit canonicalization is a deterministic post-process of the gold \texttt{line\_diff} tape: trim unchanged prefix/suffix lines, snap remaining character matches to complete old lines, and convert leftover empty old spans into pure insertions, under the hard check $\mathrm{apply}(x,\text{canonicalized})=y$.
Adjacent hunks are not merged to reduce span count; mid-line token holes are out of scope.
An example of the intended rewrite is keeping an unchanged \texttt{return output} line in the scaffold and inserting only the new function, rather than asking Stage~2 to regenerate both inside one replacement hole.
The D / D$'$ / D$''$ comparison is Table~\ref{tab:scaffold_construction}; the EOF split is Table~\ref{tab:scaffold_glue_split}.

\paragraph{Block-decision length harness.}
Each hole is seeded with a fixed-size block of \masktok{} tokens; the model denoises that block; a stop/continue decision then either appends another block or ends the hole.
Block-decision uses a dedicated control slot per block (\expandtok{} to continue, a delete/EOS token to stop); content tokens in a short final block may still resolve to EOS and trim slack.
Information still flows across the whole sequence, including prefix, suffix, sibling holes, and previously committed blocks.
DreamOn-style per-token \expandtok{}/EOS inside every mask slot is a finer-grained alternative that we do not use as the production decoder for locate-then-infill.
We likewise do not adopt FlexMDM-style insertion expectations or PILL-style length probing as Stage~2 decoders; stronger length allocation could raise the gold-scaffold ceiling without repairing locator under-coverage.
Stage~1 and Stage~2 may be separate checkpoints or a shared multitask network.

\section{Additional Result Tables}
\label{app:extra_tables}

\subsection{Matched Stage~1 Backbone Check}
\label{app:external_locators}

We fine-tune DiffuCoder, iLLaDA-8B-Instruct, and LLaDA-8B-Instruct as Stage~1 locators on the same 21,661 EditPackFT instruction, numbered-source, and gold-plan examples, with 500 validation examples. The iLLaDA and LLaDA locators use one epoch of LoRA fine-tuning; all three reported locators use checkpoint step~1354. The numbered old source precedes the noised plan answer span. We decode a fixed 160-token first plan block; DiffuCoder additionally uses its required logit right shift. Earlier iLLaDA and LLaDA checkpoints were not rescored with this decoder. On validation, 483/500 iLLaDA and 493/500 LLaDA plans are 158 tokens after stripping EOS, so these runs do not establish accurate plan-length stopping. On CanItEdit, predicted plans supply the holes; neither gold replacement text nor plan prose enters the filler prompt. All three rows use the unchanged DiffuCoder old-visible filler at step~1610 with block size~16 and confidence~$0.3$. The earlier DiffuCoder locator elsewhere in this paper places the clean numbered source after the noised plan and uses 32-token block-decision decoding; its score is not part of this matched-layout comparison.

\begin{table}[h]
\centering
\small
\setlength{\tabcolsep}{4pt}
\caption{Matched-layout Stage~1 locator outputs and CanItEdit pass@1 with the fixed DiffuCoder old-visible Stage~2 filler ($N{=}210$). Parse and exact line-operation agreement are percentages; predicted/gold edits are means.}
\label{tab:external_locators}
\begin{tabular}{lrrrrr}
  \toprule
  \textbf{Locator} & \textbf{Parse} & \textbf{Exact} & \textbf{Pred./gold} & \textbf{pass@1} & \textbf{Multi} \\
  \midrule
  DiffuCoder & 91.0 & 2.4 & 1.89 / 2.78 & 13.3 & 12.0 \\
  iLLaDA-8B & 94.3 & 4.3 & 1.93 / 2.78 & 13.8 & 11.3 \\
  LLaDA-8B & 86.7 & 3.3 & 1.81 / 2.78 & 11.9 & 10.6 \\
  \bottomrule
\end{tabular}
\end{table}

The DiffuCoder pairing reaches 16.2\% single-span, 12.4\% descriptive, and 14.3\% lazy pass@1, compared with 19.1\%, 17.1\%, and 10.5\% for iLLaDA and 14.7\%, 14.3\%, and 9.5\% for LLaDA. DiffuCoder's exact anchor agreement is 52.4\%; 23.8\% of its outputs are classified as \texttt{SyntaxError} and 61.9\% as \texttt{Exception}. These diagnostics accompany a 2.4\% exact line-operation rate: parseable plans and partly correct anchors rarely specify the complete edit operation.

\begin{table}[h]
\centering
\small
\caption{Recipe confounds in Class~A.
Locate Stage~2 curriculum is part of the two-stage interface.
Because these knobs are unmatched, Class~A cells are within-family operating points rather than a causal ranking.}
\label{tab:recipe-confounds}
\begin{tabular}{lccc}
  \toprule
  \textbf{Knob} & \textbf{Rewrite} & \textbf{\texttt{str\_replace}} & \textbf{Locate} \\
  \midrule
  Init & Instruct & Instruct & Stage-2 $1610$ \\
  LR & $2{\times}10^{-5}$ & $1{\times}10^{-5}$ & $1{\times}10^{-5}$ \\
  Batch / epochs & 128 / 4 & 64 / 5 & 64 / 3 \\
  Decode temp & $1.0$ & $0.0$ & two-stage \\
  \bottomrule
\end{tabular}
\end{table}

\begin{table}[h]
\centering
\small
\setlength{\tabcolsep}{3.5pt}
\caption{Class~A length-harness and serialization controls omitted from Table~\ref{tab:interface-controlled} (CanItEdit).
Rates are percentages.}
\label{tab:harness-controls}
\begin{tabular}{llrrrr}
  \toprule
  \textbf{Interface} & \textbf{Setup} & \textbf{pass@1} & \textbf{SynErr} & \textbf{1-span} & \textbf{multi} \\
  \midrule
  Rewrite & SFT, 256 steps & 9.0 & 36.2 & 20.6 & 3.5 \\
  Rewrite & SFT, 768 steps & 10.0 & 13.8 & 20.6 & 4.9 \\
  \texttt{str\_replace} & 256-slot, step 1515 & 9.0 & 5.7 & 23.5 & 2.1 \\
  \texttt{str\_replace} & random-pad, step 1685 & 4.3 & 11.9 & 7.4 & 2.8 \\
  Unified diff & DiffuCoder FT & 4.3 & 1.0 & 13.2 & 0.0 \\
  \bottomrule
\end{tabular}
\end{table}

\begin{table}[h]
\centering
\small
\setlength{\tabcolsep}{3.5pt}
\caption{Local vs.\ large-edit slices of CanItEdit.
Class~A cells except the last column (gold-location $+$ Qwen, Class~C).
Length similar is $|\Delta|{}\le{}5$ source lines; length grow is $\Delta\ge 6$; the remaining 2/210 shrink ($\Delta\le -6$).
Changed-line ratio ${<}10\%$ is not the same slice as 1--5 changed lines: a long file with a small ratio can still be a large absolute edit.}
\label{tab:sparse-regime}
\begin{tabular}{lrrrrr}
  \toprule
  \textbf{Bucket} & \textbf{$n$} & \texttt{str\_replace} & \textbf{Locate} & \textbf{Rewrite} & \textbf{Gold loc$+$Qwen} \\
  \midrule
  1--5 changed lines & 54 & 46.3 & 29.6 & 29.6 & 77.8 \\
  6--20 changed lines & 112 & 20.5 & 9.8 & 7.1 & 39.3 \\
  ${>}20$ changed lines & 44 & 2.3 & 0.0 & 0.0 & 9.1 \\
  Changed-line ratio ${<}10\%$ & 28 & 21.4 & 21.4 & 10.7 & 89.3 \\
  Length similar & 108 & 34.3 & 20.4 & 21.3 & 62.0 \\
  Length grow & 100 & 12.0 & 5.0 & 1.0 & 22.0 \\
  \bottomrule
\end{tabular}
\end{table}

\begin{table}[h]
\centering
\small
\caption{Existing-checkpoint train and decode curves (CanItEdit).
Rewrite 428/856 are a sibling batch-96 run; completions were chatty and fenced, then stripped before scoring.
Block-append \strreplace{} has only step 1685.
Do not plot 428--640--856 as one curve.}
\label{tab:train-curves}
\begin{tabular}{llr}
  \toprule
  \textbf{Interface} & \textbf{Checkpoints} & \textbf{pass@1 (\%)} \\
  \midrule
  Locate same-ckpt & 645 / 1000 / 1290 / 1935 & 6.7 / 11.4 / \textbf{13.3} / 12.9 \\
  Rewrite 640 decode & 256 / 512 / 768 steps & 9.0 / \textbf{11.4} / 10.0 \\
  Rewrite sibling 0522 & 428 / 856 (512 decode) & 10.5 / 11.4 \\
  \texttt{str\_replace} 256-slot & 303 / 606 / 909 / 1212 / 1515 & \textbf{11.9} / 8.1 / 8.1 / 7.6 / 9.0 \\
  \bottomrule
\end{tabular}
\end{table}

\begin{table}[h]
\centering
\small
\caption{Autoregressive rewrite references on CanItEdit, \textbf{105 descriptive instructions only}.
These cells are a different protocol from the full-evaluation tables and are not headline parity with \dllm{} results.
Rates are percentages.}
\label{tab:ar-descriptive}
\begin{tabular}{lrrrr}
  \toprule
  \textbf{Model} & \textbf{pass@1} & \textbf{SynErr} & \textbf{1-span} & \textbf{multi} \\
  \midrule
  GPT-4o & 72.4 & 0 & 79.4 & 69.0 \\
  Qwen3-8B & 51.4 & 0 & 67.6 & 43.7 \\
  Qwen2.5-Coder-7B-Instruct & 48.6 & 0 & 58.8 & 43.7 \\
  \bottomrule
\end{tabular}
\end{table}

\begin{table}[h]
\centering
\small
\caption{Coarse failure mix on CanItEdit (counts).
\texttt{AssertionError} is among scored programs, not a proof of ``right location, wrong body.''
\strreplace{} parse/apply failures are not SyntaxError.}
\label{tab:error_taxonomy}
\begin{tabular}{lrrrrr}
  \toprule
  \textbf{Interface} & \textbf{OK} & \textbf{SynErr} & \textbf{Assert.} & \textbf{Other} & \textbf{Timeout} \\
  \midrule
  Stable-DiffCoder rewrite & 127 & 0 & 56 & 26 & 1 \\
  \texttt{str\_replace} block-append & 49 & 12 & 71 & 73 & 5 \\
  Locate+infill same ckpt & 27 & 32 & 66 & 81 & 4 \\
  Rewrite SFT, 512 steps & 24 & 50 & 37 & 96 & 3 \\
  Oracle loc $+$ empty plan & 52 & 57 & 45 & 56 & 0 \\
  Oracle loc $+$ Qwen CoT & 90 & 45 & 15 & 59 & 1 \\
  \bottomrule
\end{tabular}
\end{table}

\begin{figure}[h]
\centering
\includegraphics[width=\linewidth]{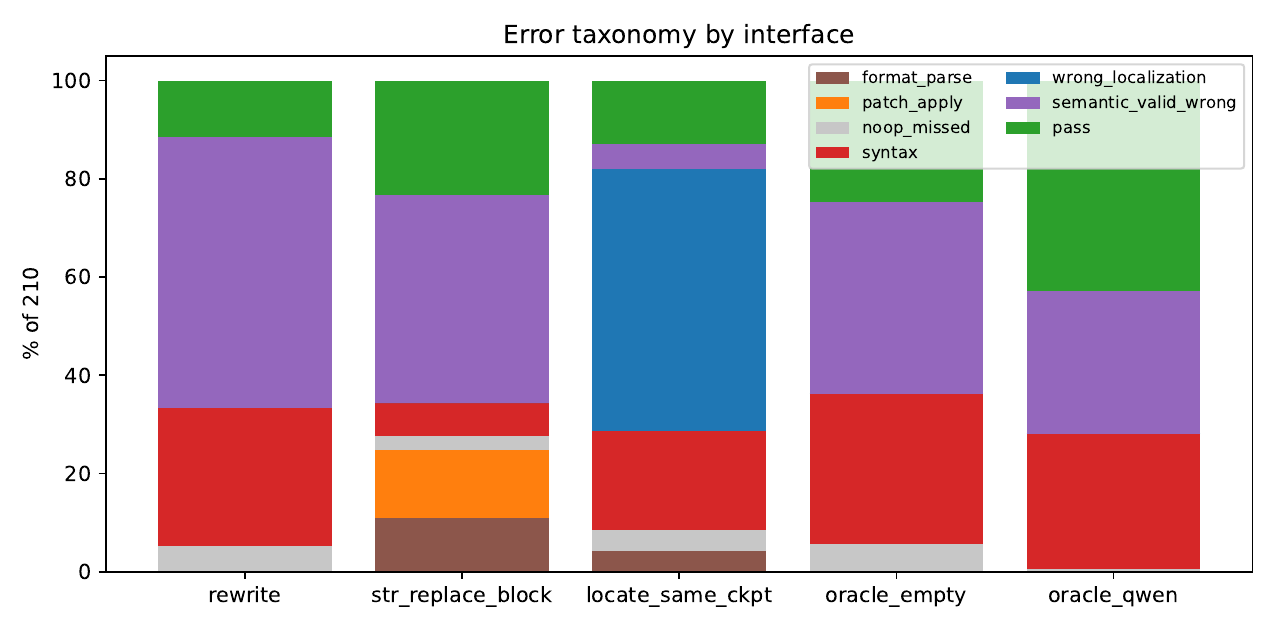}
\caption{Primary error class by interface (first-match cascade, percentages).
\texttt{wrong\_localization} is defined only for predicted-plan locate-then-infill.
IndentationError is a harness Exception and is folded into syntax.}
\label{fig:taxonomy}
\end{figure}

\begin{table}[h]
\centering
\small
\setlength{\tabcolsep}{3.5pt}
\caption{Primary error cascade on CanItEdit (counts).
First match; not a manual diagnosis.
Locate \texttt{wrong loc.} is undefined on gold-scaffold oracles.}
\label{tab:error_cascade}
\begin{tabular}{lrrrrrrr}
  \toprule
  \textbf{Interface} & \textbf{Fmt} & \textbf{Apply} & \textbf{No-op} & \textbf{Syn} & \textbf{Loc} & \textbf{Sem} & \textbf{Pass} \\
  \midrule
  Rewrite SFT, 512 steps     & 0  & 0  & 11 & 59 & 0   & 116 & 24 \\
  \strreplace{} block-append & 23 & 29 & 6  & 14 & 0   & 89  & 49 \\
  Locate+infill same ckpt    & 9  & 0  & 9  & 42 & 112 & 11  & 27 \\
  Oracle loc $+$ empty plan  & 0  & 0  & 12 & 64 & 0   & 82  & 52 \\
  Oracle loc $+$ Qwen CoT    & 0  & 0  & 1  & 58 & 0   & 61  & 90 \\
  \bottomrule
\end{tabular}
\end{table}

Locate-then-infill failures concentrate in wrong localization (112/210) rather than unparseable plans (9/210).
\strreplace{} still loses 89/210 to semantic or runtime failures after parse/apply, on top of 23 format and 29 apply errors.
A prior empty-plan audit of the multitask-1935 24.8\% cell found that 22/210 outputs dump Stage-1 prose (\texttt{Overall plan:} / \texttt{Edit N:}) into Python when the scaffold has no kept prefix before the first hole; those are format leaks of the multitask checkpoint, not near-miss infills.

\section{Additional Locate-then-Infill Diagnostics}
\label{app:diag_tables}

\begin{table}[h]
\centering
\small
\caption{Localization audits for autonomous structure predictors.}
\begin{tabular}{lr}
  \toprule
  \multicolumn{2}{c}{\strreplace{} \texttt{<old>} spans, 142 multi-span cases} \\
  \midrule
  Mean gold-span recall & 42.7\% \\
  All gold spans covered & 21.8\% (31/142) \\
  Predicted $K{=}1$ & 45/142 \\
  Enter hybrid Stage~2 branch & 53/142 \\
  \bottomrule
\end{tabular}
\label{tab:str_replace_audit}
\end{table}

\begin{table}[h]
\centering
\small
\caption{Location quality of predicted Stage~1 plans on CanItEdit.}
\begin{tabular}{lr}
  \toprule
  \textbf{Metric} & \textbf{Value} \\
  \midrule
  Parseable plans & 95.7\% \\
  Zero predicted edits & 4.3\% \\
  Exact line-op match & 13.3\% \\
  Nearest-anchor exact & 57.9\% \\
  Mean predicted / gold edits & 1.55 / 2.78 \\
  \bottomrule
\end{tabular}
\label{tab:location_quality}
\end{table}

Mean boundary distance in Table~\ref{tab:boundary_sensitivity} is extra plus missing deleted lines plus insert-anchor MAE, computed from the same \texttt{before}/\texttt{after} pairs.
Expand modes round-trip extra kept lines into the hole payload; shifts and contract do not.
These cells use gold-aware Qwen plans and the multitask filler; they are not the aligned old-visible locator-quality curve in Section~\ref{sec:scaffold_corruption}.

\begin{figure}[h]
\centering
\includegraphics[width=\linewidth]{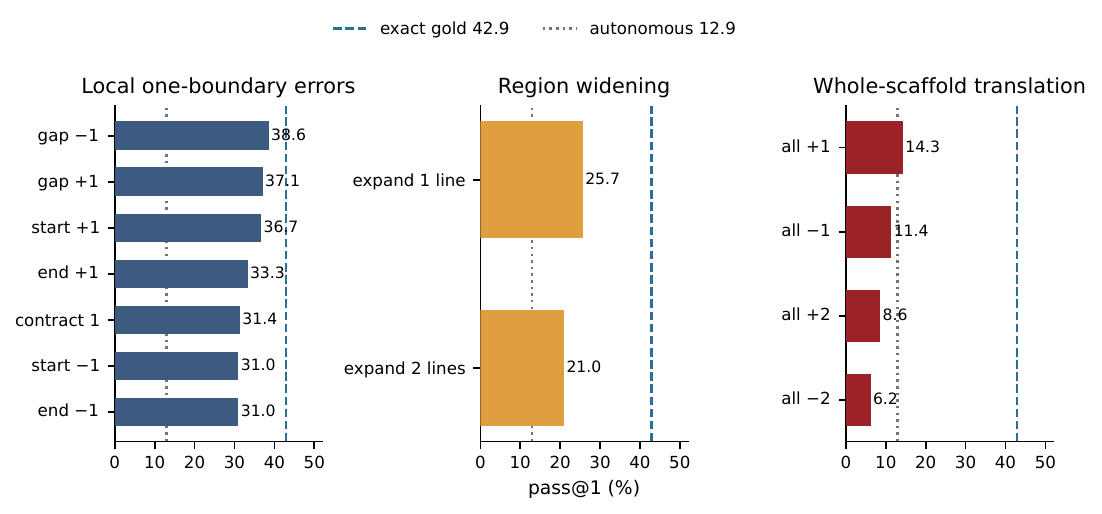}
\caption{Gold-scaffold Stage-2 pass@1 by perturbation family (shared multitask filler, gold-aware Qwen plans).
Local one-boundary errors are often survivable; expanding each range by kept lines hurts more; translating the whole scaffold is catastrophic.
Dashed/dotted lines: exact-gold privileged fill 42.9\% and autonomous same-checkpoint 12.9\% (references, not a ranking).
Diagnostic, not autonomous performance.}
\label{fig:boundary}
\end{figure}

\begin{table}[h]
\centering
\small
\setlength{\tabcolsep}{3.5pt}
\caption{Gold-scaffold Stage-2 fill under boundary perturbations (multitask step~1935 filler, gold-aware Qwen plan).
Diagnostic, not autonomous performance.}
\label{tab:boundary_sensitivity}
\resizebox{\linewidth}{!}{%
\begin{tabular}{lrrrrr}
  \toprule
  \textbf{Scaffold} & \textbf{Dist.} & \textbf{pass@1} & \textbf{SynErr} & \textbf{1-span} & \textbf{Multi} \\
  \midrule
  Exact gold ranges                         & 0.00 & 42.9\% & 21.4\% & 57.4\% & 35.9\% \\
  \midrule
  \multicolumn{6}{l}{\textit{Local typed $\pm 1$}} \\
  Shift insert-gap $-1$                     & 0.69 & 38.6\% & 21.0\% & 50.0\% & 33.1\% \\
  Shift insert-gap $+1$                     & 0.71 & 37.1\% & 22.9\% & 50.0\% & 31.0\% \\
  Shift start $+1$ (where valid)            & 0.56 & 36.7\% & 21.4\% & 54.4\% & 28.2\% \\
  Shift end $+1$                            & 1.66 & 33.3\% & 31.4\% & 55.9\% & 22.5\% \\
  Contract each range by 1 (where valid)    & 0.83 & 31.4\% & 23.3\% & 47.1\% & 23.9\% \\
  Shift start $-1$                          & 1.98 & 31.0\% & 29.5\% & 48.5\% & 22.5\% \\
  Shift end $-1$ (where valid)              & 0.56 & 31.0\% & 21.0\% & 47.1\% & 23.2\% \\
  \midrule
  \multicolumn{6}{l}{\textit{Expand (extra kept lines enter the hole)}} \\
  Expand each range by 1 kept line          & 5.15 & 25.7\% & 38.1\% & 47.1\% & 15.5\% \\
  Expand each range by 2 kept lines         & 8.37 & 21.0\% & 33.3\% & 42.6\% & 10.6\% \\
  \midrule
  \multicolumn{6}{l}{\textit{Rigid translation of every hole}} \\
  Shift every range/gap by $+1$             & 5.79 & 14.3\% & 20.0\% & 27.9\% & 7.7\% \\
  Shift every range/gap by $-1$             & 6.73 & 11.4\% & 10.5\% & 16.2\% & 9.2\% \\
  Shift every range/gap by $+2$             & 9.38 & 8.6\%  & 21.9\% & 16.2\% & 4.9\% \\
  Shift every range/gap by $-2$             & 10.93 & 6.2\% & 11.4\% & 8.8\% & 4.9\% \\
  \bottomrule
\end{tabular}%
}
\end{table}

\subsection{Conditioning Text and Oracle Scope}
\label{sec:rationale}

\begin{table}[h]
\centering
\caption{Gold-location Stage-2 fill under different plan texts.
Rows above the last midrule use the multitask step~1935 filler unless noted (ep5).
The last two rows are separately trained aligned Stage-2 diagnostics from DiffuCoder-Instruct (block~16, five epochs): \texttt{empty\_plan} and \texttt{original\_source} at \texttt{global\_step\_1610}, not the plan-conditioned ep5 specialist.
Only that pair is a matched test of old-source visibility; 20.5/27.1 vs.\ 24.8 is not.
The integrated-diff pre-glue $20.5\%$ cell in Table~\ref{tab:scaffold_construction} is a third, different $20.5\%$.
All rows are diagnostics, not autonomous performance.}
\label{tab:oracle_conditioning}
\small
\begin{tabular}{lrrrr}
  \toprule
  \textbf{Plan / conditioning} & \textbf{pass@1} & \textbf{1-span} & \textbf{Multi} & \textbf{SynErr} \\
  \midrule
  Gold loc.\ + Qwen CoT                         & 42.9\% & 57.4\% & 35.9\% & 21.4\% \\
  Gold loc.\ + location-free gold-aware Qwen    & 42.9\% & 57.4\% & 35.9\% & 22.4\% \\
  Gold loc.\ + Qwen CoT (ep5 filler)            & 39.5\% & 48.5\% & 35.2\% & 27.1\% \\
  Gold loc.\ + empty plan (multitask 1935)      & 24.8\% & 42.6\% & 16.2\% & 27.1\% \\
  Gold loc.\ + input-only Qwen3-8B              & 24.8\% & 41.2\% & 16.9\% & 21.4\% \\
  Gold loc.\ + Qwen headers only                & 22.4\% & 42.6\% & 12.7\% & 16.7\% \\
  Gold loc.\ + model CoT                        & 21.0\% & 36.8\% & 13.4\% & 19.5\% \\
  Gold loc.\ + gold headers + model rationale   & 20.0\% & 36.8\% & 12.0\% & 20.5\% \\
  \midrule
  \multicolumn{5}{l}{\textit{Aligned Stage-2 SFT (DiffuCoder-Instruct, step 1610)}} \\
  Gold loc.\ + aligned empty plan               & 20.5\% & 33.8\% & 14.1\% & 26.7\% \\
  Gold loc.\ + aligned old-visible context      & 27.1\% & 30.9\% & 25.4\% & 27.1\% \\
  \bottomrule
\end{tabular}
\end{table}

\subsection{Performance by Number of Holes}
\label{sec:multispan}

\begin{table}[h]
\centering
\small
\caption{Pass@1 by gold insert-span count (CanItEdit).
Gold-location rows give Stage~2 every gold hole (empty-plan row: multitask 1935 filler).
Class~A rows are autonomous.
$4+$ pools 4--10 holes ($n{=}60$).}
\label{tab:hole_count}
\begin{tabular}{lrrrr}
  \toprule
  \textbf{Interface} & \textbf{1 hole} & \textbf{2} & \textbf{3} & \textbf{$4+$} \\
  \midrule
  $n$ & 68 & 54 & 28 & 60 \\
  \midrule
  Oracle loc.\ $+$ Qwen CoT (C) & 57.4\% & 50.0\% & 50.0\% & 16.7\% \\
  Oracle loc.\ $+$ empty (C)    & 42.6\% & 22.2\% & 21.4\% & 8.3\% \\
  \strreplace{} block-append (A) & 48.5\% & 16.7\% & 17.9\% & 3.3\% \\
  Locate-and-infill same ckpt (A) & 27.9\% & 9.3\% & 7.1\% & 1.7\% \\
  Rewrite SFT 512 (A)           & 27.9\% & 7.4\% & 3.6\% & 0.0\% \\
  \bottomrule
\end{tabular}
\end{table}

\begin{table}[h]
\centering
\small
\caption{Pass@1 / SyntaxError by gold insert-span count (CanItEdit).
SyntaxError is not monotonic in hole count.
The old-visible 1-hole cell has the highest SyntaxError among 1--2 holes; privileged Qwen
CoT SyntaxError is nearly flat through $4+$.
Diagnostic, not autonomous performance.}
\label{tab:hole_syntax}
\begin{tabular}{lrrrr}
  \toprule
  \textbf{Interface} & \textbf{1 hole} & \textbf{2} & \textbf{3} & \textbf{$4+$} \\
  \midrule
  $n$ & 68 & 54 & 28 & 60 \\
  \midrule
  Oracle loc.\ $+$ Qwen CoT (1935) & 57.4 / 20.6 & 50.0 / 22.2 & 50.0 / 21.4 & 16.7 / 21.7 \\
  Oracle loc.\ $+$ empty (1935)    & 42.6 / 16.2 & 22.2 / 25.9 & 21.4 / 28.6 & 8.3 / 40.0 \\
  Aligned empty 1610               & 33.8 / 27.9 & 27.8 / 20.4 & 3.6 / 17.9 & 6.7 / 35.0 \\
  Aligned old-visible 1610         & 30.9 / 35.3 & 44.4 / 11.1 & 7.1 / 28.6 & 16.7 / 31.7 \\
  \bottomrule
\end{tabular}
\end{table}

\begin{table}[h]
\centering
\small
\caption{One-span interventions on gold scaffolds with gold-aware Qwen plans (filler 1935).
Single-span drop is 0\% by construction.
Diagnostic, not autonomous performance.}
\label{tab:onespan_interv}
\begin{tabular}{lrrr}
  \toprule
  \textbf{Intervention} & \textbf{pass@1} & \textbf{1-span} & \textbf{Multi} \\
  \midrule
  Exact gold spans                         & 42.9\% & 57.4\% & 35.9\% \\
  Extra false-positive span                & 39.5\% & 48.5\% & 35.2\% \\
  Shift first span $+1$ line               & 19.5\% & 29.4\% & 14.8\% \\
  Drop last gold span                      & 5.7\%  & 0.0\%  & 8.5\% \\
  \bottomrule
\end{tabular}
\end{table}

\begin{table}[h]
\centering
\small
\caption{Gold edit-count prompt hint on Stage~1 (multitask 1935, confidence 0.1) filled by the same checkpoint.
Raises exact-count agreement, not pass@1.
Diagnostic, not a production decoder.}
\label{tab:goldk}
\begin{tabular}{lrrrr}
  \toprule
  \textbf{Planner} & \textbf{Parse} & \textbf{Exact op} & \textbf{Mean pred} & \textbf{pass@1} \\
  \midrule
  Unconstrained Stage~1 & 95.7\% & 13.3\% & 1.55 & 12.9\% \\
  Gold-$K$ prompt hint  & 89.0\% & 14.3\% & 1.80 & 11.9\% \\
  \bottomrule
\end{tabular}
\end{table}

\section{Integrated Gold-Scaffold Construction}
\label{app:scaffold_construction}
\label{sec:scaffold_construction}

This diagnostic holds the aligned integrated-diff filler at
\texttt{global\_step\_1610} fixed (block~16, 64 steps, confidence $0.3$, no plan file)
and changes only gold-scaffold construction.
D / D$'$ / D$''$ are therefore not three checkpoints.
They are also not the aligned empty-plan $20.5\%$ cell in Table~\ref{tab:oracle_conditioning}.

\paragraph{Pre-glue artifact.}
Every CanItEdit source lacks a final newline ($105/105$).
The original serializer could glue the terminal old line onto the following mask as
\texttt{-OLD+[MASK]}, so gold reconstruction failed on $43/105$ programs ($86$
descriptive$+$lazy evaluations).
That slice scored $0\%$ pass@1.
Keep-line audits of the reconstructions do not support reading this as the filler
overwriting immutable keep tokens.

\paragraph{Glue-fix.}
Minus records are newline-terminated so that
$\mathrm{reconstruct}(\text{gold canvas})=y$ on all $105$ sources.
Unaffected evaluations are identical at $34.7\%$; the whole $20.5{\to}26.2$ gain is the
affected slice going $0.0\% \to 14.0\%$.

\paragraph{Minimal-edit canonicalization.}
After glue-fix, each gold hunk is trimmed so unchanged prefix/suffix lines remain in the
scaffold.
A schematic non-minimal replacement
\begin{lstlisting}
-    return output
+    return output
+
+def header(...):
\end{lstlisting}
becomes a keep plus a pure insertion of the new function.
The hard check is $\mathrm{apply}(x,\text{canonicalized gold edits})=y$.
Span count is unchanged (mean $2.78$); $18/43$ EOF/no-newline replacements become
inserts.
The further $+1.9$ pass@1 ($55{\to}59$ OK, paired $+4/-0$) is confined to that
$43$-source slice ($14.0\% \to 18.6\%$).
Both wins are \texttt{12\_linkedlist\_sort} and \texttt{63\_knary\_trees} (both prompts).
\texttt{10\_csv\_parser} still fails, so tighter scaffolding is not sufficient.

\begin{table}[h]
\centering
\small
\setlength{\tabcolsep}{3.2pt}
\caption{EOF/no-newline glue split on the integrated-diff step-1610 filler (CanItEdit).
Affected: $86$ evaluations / $43$ sources.
Unaffected: $124$ evaluations / $62$ sources.
Oracle-length locks gold occupancy on the glue-fixed scaffold (train--test mismatch).
Rates are percentages.
Diagnostic, not autonomous performance.}
\label{tab:scaffold_glue_split}
\begin{tabular}{lrrr}
  \toprule
  \textbf{Arm} & \textbf{Full} & \textbf{Unaffected} & \textbf{Affected} \\
  \midrule
  D pre-glue & 20.5 & 34.7 & 0.0 \\
  D$'$ glue-fix & 26.2 & 34.7 & 14.0 \\
  D$''$ canonicalize & 28.1 & 34.7 & 18.6 \\
  Glue-fix $+$ oracle length & 27.1 & 30.6 & 22.1 \\
  \bottomrule
\end{tabular}
\end{table}

\begin{table}[h]
\centering
\small
\caption{Gold-scaffold construction with oracle locations. D, D$'$, and D$''$ share one integrated-diff step-1610 filler; only scaffold construction changes.}
\label{tab:scaffold_construction}
\begin{tabular}{lrrrr}
  \toprule
  \textbf{Arm} & \textbf{pass@1} & \textbf{1-span} & \textbf{Multi} & \textbf{SynErr} \\
  \midrule
  D pre-glue integrated & 20.5 & 30.9 & 15.5 & 20.5 \\
  D$'$ glue-fixed integrated & 26.2 & 33.8 & 22.5 & 26.7 \\
  D$''$ canonicalized integrated & 28.1 & 33.8 & 25.4 & 24.8 \\
  Local $\pm2$ prefix control & 28.1 & 38.2 & 23.2 & 21.9 \\
  \bottomrule
\end{tabular}
\end{table}

Inference-time oracle length is de-emphasized: the checkpoint was trained for
variable-length block generation, so gold occupancy adds both extra oracle information
and a decode mismatch.
It scores below canonicalized variable-length $28.1\%$, raises SyntaxError to $40.0\%$,
and degrades the unaffected slice.
We do not treat $27.1\%$ as a Stage-2 upper bound or as evidence that length control is
unimportant.

\section{Composition Diagnostic: Predicted \texttt{<old>} $\to$ Old-Visible Infill}
\label{app:composition}

The hybrid in Section~\ref{sec:composition} is inference-only.
It reuses the block-append \strreplace{} dump at step~1685 as a locator and the aligned old-visible Stage~2 checkpoint
(\path{stage3_oracle_scaffold_diffucoder_original_source/global_step_1610})
as a filler.
Decode matches that filler: block size 16, decision slot, 64 steps, confidence $0.3$, \texttt{original\_source} context, no plan file.

\paragraph{Routing.}
Hybrid-multi keeps the original \strreplace{} reconstruction when the model predicts $K{=}1$.
When $K{\ge}2$ and every \texttt{<old>} span is an exact unique substring of the source, those spans become Stage~2 holes and the generated \texttt{<new>} text is discarded.
Any missing, empty, or non-unique \texttt{<old>} copies the \strreplace{} reconstruction instead of forcing an infill.
Gold locations are never used.
Hybrid-all routes every uniquely matched \texttt{<old>} through Stage~2, including $K{=}1$; we report it only as a control, not as a method.

\paragraph{Matched repairs of predicted holes.}
Figure~\ref{fig:main-method}'s repair ladder uses a separate predicted-hole baseline $P$, not Hybrid-multi. For every example, $P$ turns uniquely matched \texttt{<old>} spans from the step-1685 dump into holes, merging overlapping predicted ranges; missing or ambiguous matches are dropped, and a parse failure leaves the source unchanged. Predicted \texttt{<new>} text is never used, and there is no \strreplace{} fallback. Coverage repair ($C$) adds each gold span that overlaps no predicted hole without changing predicted boundaries. Boundary repair ($B$) replaces the predicted geometry in each predicted--gold overlap component with the exact gold span or spans; unmatched predicted holes remain, and missing gold spans are not added. Applying both repairs therefore retains unmatched false-positive holes and need not produce the gold scaffold. All four tapes use the same old-visible step-1610 filler and decoding settings as the 25.4\% gold cell, with no plan file.

\begin{table}[h]
\centering
\small
\caption{Paired oracle repairs of the separate predicted-hole baseline $P$ on 142 multi-span CanItEdit examples. The gold row is the existing clean-scaffold result. Repairs change scaffold geometry, not the filler or decoder; none is an autonomous editor.}
\label{tab:repair_ladder}
\begin{tabular}{lrr}
  \toprule
  \textbf{Scaffold} & \textbf{Passed} & \textbf{pass@1 (\%)} \\
  \midrule
  Predicted baseline $P$ & 15/142 & 10.6 \\
  $P+C$ (coverage) & 19/142 & 13.4 \\
  $P+B$ (boundaries) & 20/142 & 14.1 \\
  $P+C+B$ (both) & 31/142 & 21.8 \\
  Gold & 36/142 & 25.4 \\
  \bottomrule
\end{tabular}
\end{table}

On 94/142 examples, $P+C+B$ has exactly the gold span list. The remaining 48 retain at least one unmatched predicted hole. Relative to $P$, the paired 95\% bootstrap interval for $P+C+B$ is $+4.9$ to $+17.6$ pass@1 points; the corresponding intervals for $P+C$ and $P+B$ include zero. The individual repairs therefore show directional gains, while their combined improvement is better supported by the paired comparison.

\begin{table}[h]
\centering
\small
\caption{Detailed composition cells (CanItEdit).
Hybrid-multi overall 22.9\% is not a better editor: single-span 48.5\% is inherited from direct \strreplace{}.
Rates are percentages.}
\label{tab:composition_detail}
\begin{tabular}{p{0.20\linewidth}p{0.28\linewidth}rrrr}
  \toprule
  \textbf{Interface} & \textbf{Setup} & \textbf{pass@1} & \textbf{SynErr} & \textbf{1-span} & \textbf{Multi} \\
  \midrule
  Direct \texttt{str\_replace} & block-append 1685 & 23.3 & 5.7 & 48.5 & 11.3 \\
  Hybrid-multi & predicted \texttt{<old>} $\to$ old-visible & 22.9 & 9.0 & 48.5 & 10.6 \\
  Hybrid-all & all unique \texttt{<old>} $\to$ old-visible & 13.3 & 23.8 & 22.1 & 9.2 \\
  Gold loc.\ $+$ old-visible & aligned 1610 & 27.1 & 27.1 & 30.9 & 25.4 \\
  Gold loc.\ $+$ aligned empty & aligned 1610 & 20.5 & 26.7 & 33.8 & 14.1 \\
  \bottomrule
\end{tabular}
\end{table}

\paragraph{Fully-covered subset.}
On the 31 multi-span examples whose predicted \texttt{<old>} spans cover every gold hunk, gold-location old-visible Stage~2 on the same filler is $13/31$ (Table~\ref{tab:paired31}).
Hybrid-multi actually routed only $19/31$ of these examples to Stage~2; the other $12$ keep the direct \strreplace{} reconstruction because predicted $K{=}1$.
McNemar discordant pairs on the 31 IDs favor gold over predicted-scaffold Stage~2 ($n_{10}{=}5$, $n_{01}{=}0$), but example-resample 95\% CIs overlap ($25.8$--$58.1\%$ vs.\ $9.7$--$41.9\%$).
We therefore do not treat $13/31$ vs.\ $8/31$ as a second major bottleneck; it is supporting evidence that coverage is not sufficient.
A deterministic boundary-relation breakdown (first match: merge, fragment, exact, strict superset, strict subset, residual overlap) is in Table~\ref{tab:paired31_boundary}; several bins have $n{\le}3$ and are not interpreted.

\begin{table}[h]
\centering
\small
\caption{Paired 31-case control on multi-span examples whose predicted \texttt{<old>} spans cover every gold hunk.
Filler is aligned old-visible Stage~2 at step~1610.
Routed $=$ Hybrid-multi sent the example to Stage~2; fallback $=$ predicted $K{=}1$ kept the \strreplace{} reconstruction.}
\label{tab:paired31}
\begin{tabular}{lrrr}
  \toprule
  \textbf{Method} & \textbf{All 31} & \textbf{Routed 19} & \textbf{Fallback 12} \\
  \midrule
  Direct \texttt{str\_replace} & 10/31 & 7/19 & 3/12 \\
  Predicted scaffold $\to$ filler & 8/31 & 5/19 & 3/12 \\
  Gold scaffold $\to$ same filler & 13/31 & 7/19 & 6/12 \\
  \bottomrule
\end{tabular}
\end{table}

\begin{table}[h]
\centering
\small
\setlength{\tabcolsep}{3.5pt}
\caption{Boundary-relation and $K$-match slices of the 31 fully-covered examples.
Relation rows use first-match precedence (merge $>$ fragment $>$ exact $>$ superset $>$ subset $>$ residual overlap).
The $K$ rows are not exclusive of the relation rows; bins with $n{\le}3$ are not interpreted.}
\label{tab:paired31_boundary}
\begin{tabular}{lrrrr}
  \toprule
  \textbf{Relation} & \textbf{$n$} & \textbf{Direct} & \textbf{Pred.\ scaf.} & \textbf{Gold scaf.} \\
  \midrule
  Merged multiple gold spans & 12 & 3/12 & 3/12 & 6/12 \\
  Strict superset             & 10 & 5/10 & 2/10 & 3/10 \\
  Exact boundaries            & 3  & 2/3  & 1/3  & 1/3 \\
  Strict subset               & 3  & 0/3  & 2/3  & 3/3 \\
  Overlap / fragmented        & 3  & 0/3  & 0/3  & 0/3 \\
  \midrule
  Predicted $K{=}$ gold $K$   & 17 & 7/17 & 5/17 & 5/17 \\
  Predicted $K{\neq}$ gold $K$ & 14 & 3/14 & 3/14 & 8/14 \\
  \bottomrule
\end{tabular}
\end{table}

\paragraph{Full gold-scaffold corruption grid.}
Table~\ref{tab:scaffold_corruption_full} expands Table~\ref{tab:scaffold_corruption} with overall pass@1, including shrink / split / merge and the $\pm 1$ translations used in Figure~\ref{fig:locator_quality}.
Operators are deterministic; impossible corruptions leave the gold scaffold.
Split-one applies on $76/210$ problems and merge-nearby on $58/210$; those two pass@1 numbers therefore mix applied corruptions with unmodified gold cases and are not used as headline evidence.

\begin{figure}[h]
\centering
\includegraphics[width=\linewidth]{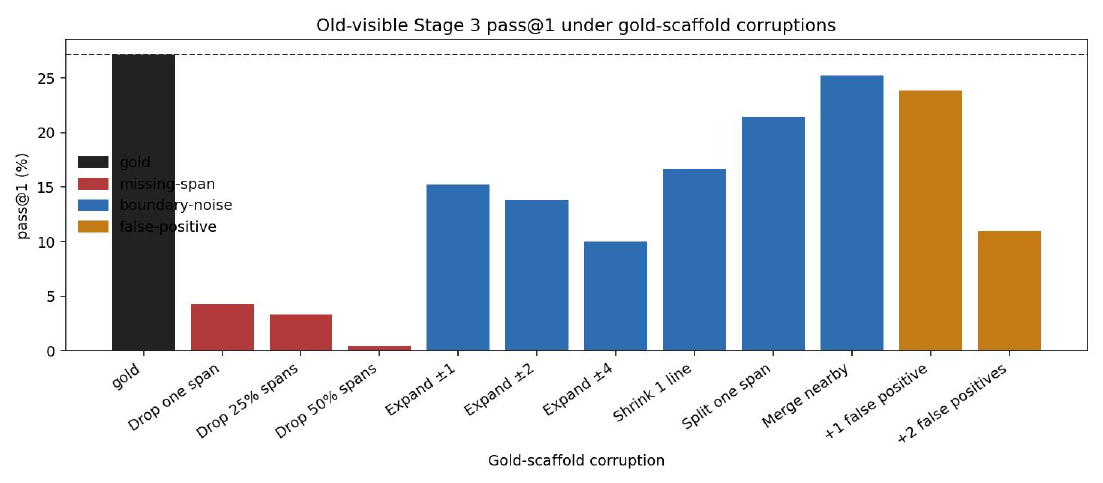}
\caption{Overall CanItEdit pass@1 under gold-scaffold corruptions of the aligned old-visible filler.
Missing required spans collapse performance; one extra hole is comparatively cheap; expanding gold boundaries is not.
Drop-one / drop-25 / drop-50 overall rates mix emptied 1-span scaffolds; the locator-quality curve in Figure~\ref{fig:locator_quality} uses multi-span pass@1.
Diagnostic, not autonomous performance.}
\label{fig:corruption}
\end{figure}

\begin{table}[h]
\centering
\small
\setlength{\tabcolsep}{3.2pt}
\caption{Full gold-scaffold corruption grid on the aligned old-visible filler (CanItEdit).
Relative $\Delta$ is against clean gold $57/210$ ($27.1\%$).
Applied / impossible counts how often the operator changed the scaffold.
Shift $\pm 1$ also changes coverage (mean recall 29.6\% / 32.9\% on the 142) and extra-span count.}
\label{tab:scaffold_corruption_full}
\begin{tabular}{lrrrrr}
  \toprule
  \textbf{Scaffold} & \textbf{pass@1} & \textbf{Multi} & \textbf{$\Delta$} & \textbf{SynErr} & \textbf{Appl./imposs.} \\
  \midrule
  Clean gold                      & 27.1 & 25.4 & ---     & 27.1 & 210/0 \\
  Drop one gold span              & 4.3  & 6.3  & $-22.9$ & 12.9 & 210/0 \\
  Drop 25\% of gold spans         & 3.3  & 4.9  & $-23.8$ & 11.0 & 210/0 \\
  Drop 50\% of gold spans         & 0.5  & 0.7  & $-26.7$ & 10.0 & 210/0 \\
  Expand $\pm 1$                  & 15.2 & 7.7  & $-11.9$ & 39.0 & 210/0 \\
  Expand $\pm 2$                  & 13.8 & 7.0  & $-13.3$ & 42.9 & 210/0 \\
  Expand $\pm 4$                  & 10.0 & 7.0  & $-17.1$ & 40.0 & 210/0 \\
  Shrink 1 line                   & 16.7 & 11.3 & $-10.5$ & 30.0 & 210/0 \\
  Split one span                  & 21.4 & 18.3 & $-5.7$  & 36.7 & 76/134 \\
  Merge nearby                    & 25.2 & 22.5 & $-1.9$  & 29.0 & 58/152 \\
  $+1$ false-positive hole        & 23.8 & 20.4 & $-3.3$  & 30.5 & 208/2 \\
  $+2$ false-positive holes       & 11.0 & 8.5  & $-16.2$ & 46.7 & 204/6 \\
  Shift every range $-1$          & 11.9 & 9.9  & $-15.2$ & 15.2 & 210/0 \\
  Shift every range $+1$          & 8.1  & 4.9  & $-19.0$ & 32.4 & 210/0 \\
  \bottomrule
\end{tabular}
\end{table}

\paragraph{Locator-quality response (14 cells).}
Table~\ref{tab:locator_quality} is the numerical companion to Figure~\ref{fig:locator_quality}.
All corruption rows use the aligned old-visible filler with gold locations corrupted by a named operator.
The last two rows are autonomous reference operating points at the \strreplace{} locator's measured 42.7\% mean gold-span recall; they are not matched causal baselines.
Do not put the privileged 42.9\% Qwen-CoT cell on this curve.

\begin{table}[h]
\centering
\small
\setlength{\tabcolsep}{3.0pt}
\caption{Locator quality $\to$ multi-span pass@1 on the aligned old-visible filler.
Recall and extra-span count are means on the 142 multi-span examples.
$\Delta$ vs.\ gold and vs.\ 11.3 are multi-span points.
Corruption rows are diagnostics; \strreplace{} rows are autonomous references.}
\label{tab:locator_quality}
\resizebox{\linewidth}{!}{%
\begin{tabular}{llrrrrr}
  \toprule
  \textbf{Family} & \textbf{Mode} & \textbf{Recall} & \textbf{Extra} & \textbf{Multi} & \textbf{Overall} & \textbf{SynErr} \\
  \midrule
  gold & clean scaffold & 100.0 & 0.00 & 25.4 & 27.1 & 27.1 \\
  missing-span & drop-one & 66.0 & 0.00 & 6.3 & 4.3 & 12.9 \\
  missing-span & drop-25\% & 62.2 & 0.00 & 4.9 & 3.3 & 11.0 \\
  missing-span & drop-50\% & 45.5 & 0.00 & 0.7 & 0.5 & 10.0 \\
  shift & $-1$ line & 29.6 & 2.62 & 9.9 & 11.9 & 15.2 \\
  shift & $+1$ line & 32.9 & 2.46 & 4.9 & 8.1 & 32.4 \\
  \midrule
  boundary & expand-1 & 100.0 & 0.00 & 7.7 & 15.2 & 39.0 \\
  boundary & expand-2 & 100.0 & 0.00 & 7.0 & 13.8 & 42.9 \\
  boundary & expand-4 & 100.0 & 0.00 & 7.0 & 10.0 & 40.0 \\
  boundary & contract-1 & 100.0 & 0.73 & 11.3 & 16.7 & 30.0 \\
  boundary & split-one & 100.0 & 0.00 & 18.3 & 21.4 & 36.7 \\
  boundary & merge-nearby & 100.0 & 0.00 & 22.5 & 25.2 & 29.0 \\
  false-positive & $+1$ hole & 100.0 & 1.83 & 20.4 & 23.8 & 30.5 \\
  false-positive & $+2$ holes & 100.0 & 2.83 & 8.5 & 11.0 & 46.7 \\
  \midrule
  autonomous & direct \strreplace{} & 42.7 & 0.68 & 11.3 & 23.3 & --- \\
  autonomous & hybrid-multi & 42.7 & 0.68 & 10.6 & 22.9 & --- \\
  \bottomrule
\end{tabular}%
}
\end{table}

\section{Sequential \texttt{str\_replace} Control}
\label{app:iterative}
\label{sec:iterative}

This inference-only control uses the same block-append
\strreplace{} checkpoint as the 23.3\% / 48.5\% / 11.3\% cell
(\texttt{checkpoints/str\_replace\_sft\_block/global\_step\_1685}; block-append,
decision slot, block size 32).
No Stage~2 and no new training.
Truncations at max $1$--$4$ rounds are taken from the same max-$5$ trajectories.
Gold-location old-visible Stage~2 is an oracle reference, not an autonomous baseline.

\paragraph{One-edit prompt.}
The model is instructed to emit exactly one \texttt{<edit>} or \texttt{STOP} per round.
The runtime rejects any document with more than one replacement (\texttt{too\_many\_edits}).
This is a protocol mismatch: 52.9\% of trajectories (111/210) die that way, and
mean accepted edits is 1.04.

\paragraph{First-valid executor.}
The original one-shot \strreplace{} prompt is kept.
The model may serialize several replacements; the executor commits the first uniquely
applicable non-noop edit, discards the rest, and replans on the updated file.
Invalid remaining proposals terminate the trajectory.
Mean rounds executed: 3.07; mean accepted edits: 2.20; mean proposed edits per round: 1.98.

\begin{table}[h]
\centering
\small
\caption{Sequential \strreplace{} on CanItEdit.
One-edit prompt and first-valid executor use the same checkpoint as direct one-shot.
Rates are percentages.
Gold-location old-visible Stage~2 is a diagnostic, not a sequential baseline.}
\label{tab:iterative}
\begin{tabular}{lrrr}
  \toprule
  \textbf{Method} & \textbf{Overall} & \textbf{1-span} & \textbf{Multi-span} \\
  \midrule
  Direct one-shot \texttt{str\_replace} & 23.3 & 48.5 & 11.3 \\
  One-edit prompt, max 5 rounds & 14.8 & 35.3 & 4.9 \\
  \midrule
  First-valid executor, max 1 round & 19.5 & 48.5 & 5.6 \\
  First-valid executor, max 2 rounds & 18.1 & 38.2 & 8.5 \\
  First-valid executor, max 3 rounds & 17.1 & 30.9 & 10.6 \\
  First-valid executor, max 4 rounds & 19.0 & 36.8 & 10.6 \\
  First-valid executor, max 5 rounds & 19.5 & 36.8 & 11.3 \\
  \midrule
  Gold-loc.\ old-visible Stage~2 & 27.1 & 30.9 & 25.4 \\
  \bottomrule
\end{tabular}
\end{table}

\begin{table}[h]
\centering
\small
\caption{First-valid sequential stop reasons (max 5 rounds).
\texttt{too\_many\_edits} is the one-edit-prompt death mode (52.9\%) and does not apply here.}
\label{tab:iterative_stops}
\begin{tabular}{lrr}
  \toprule
  \textbf{Stop reason} & \textbf{$n$} & \textbf{Rate} \\
  \midrule
  \texttt{noop\_edit} & 59 & 28.1\% \\
  Malformed edit structure & 56 & 26.7\% \\
  \texttt{repeated\_edit} & 29 & 13.8\% \\
  \texttt{max\_rounds} & 27 & 12.9\% \\
  \texttt{code\_state\_loop} & 22 & 10.5\% \\
  \texttt{ambiguous\_old} & 13 & 6.2\% \\
  \texttt{old\_not\_found} & 4 & 1.9\% \\
  \bottomrule
\end{tabular}
\end{table}

On the 142 multi-span problems, first-valid max~5 recovers 16 successes (11.3\%), of which
14/16 used more than one accepted edit.
Round~1 damages syntax or increases remaining gold hunks on 34.5\% (49/142).
Zero trajectories stop after exactly one accepted edit.
The 48.5\%$\to$36.8\% single-span regression from max~1 to max~5 is therefore not hidden by
the multi-span recovery (5.6\%$\to$11.3\%).

\section{FineEdit Wiki Targeted-Editing Probe}
\label{app:fineedit}
We use 8,024 Wiki examples from FineEdit (\texttt{YimingZeng/FineEdit\_bench}). Each sentence becomes one numbered pseudo-line, allowing the same line-diff procedure to construct \strreplace{} operations, locator targets, and old-visible infilling scaffolds. In 84.1\% of examples, at least three sentences are edited; the median is seven edited sentences out of 37. Gold \strreplace{} operations apply on 99.98\% of pseudo-documents. Examples sharing a 12-word shingle are assigned to the same split, yielding 6,687 training, 649 development, and 688 test examples.

Separate \strreplace{}, locator, and filler models are fine-tuned from Dream-v0-Instruct-7B for two epochs. Training excludes examples exceeding 2,048 tokens; the respective retained training sets contain 2,672, 4,232, and 2,854 examples. Evaluation uses the first 64 test examples sorted by id, including examples that would not fit the training filter, and scores outputs against the sentence-level pseudo-target rather than raw WikiText. Sentence splitting changes the original WikiText representation, and reference spacing changes often widen gold regions. The locator's plans parse on 57.8\% of these examples and completely cover the gold regions on 48.4\%; it predicts 6.0 regions on average against 3.56 gold regions.

\end{document}